\documentclass[a4paper,11pt]{article}
\pdfoutput=1 

\usepackage{jheppub}

\usepackage[T1]{fontenc} 

\usepackage[utf8]{inputenc}
\usepackage[british]{babel}
\usepackage[Symbolsmallscale]{upgreek}
\usepackage{isomath}
\usepackage{xcolor}
\usepackage{graphicx}

\title{Probing the gravitational wave background from cosmic strings with LISA}

\def\nsf{\mathsf{n}}
\def\lsim{\mathrel{\rlap{\lower3pt\hbox{\hskip0pt$\sim$}}
   \raise1pt\hbox{$<$}}}         
\def\gsim{\mathrel{\rlap{\lower4pt\hbox{\hskip1pt$\sim$}}
   \raise1pt\hbox{$>$}}}         

\def\be{\begin{equation}}
\def\ee{\end{equation}}
\def\bea{\begin{eqnarray}}
\def\eea{\end{eqnarray}}

\newcommand{\cc}{{\tilde c}}
\newcommand{\al}{\alpha_L}

\newcommand{\vv}{{\bar v}}

\makeatletter
\def\blfootnote{\xdef\@thefnmark{}\@footnotetext}
\makeatother

\author[\,]{Pierre Auclair$^{\,a}$,}
\author[\,]{Jose J.\,Blanco-Pillado$^{\,b,c}$,}
\author[\,]{Daniel G. Figueroa$^{\,d,e,\dagger}$,\blfootnote{$^\dagger\,$Project coordinator: daniel.figueroa@cern.ch}}
\author[\,]{Alexander C.~Jenkins$^{\,f}$,}
\author[\,]{Marek Lewicki$^{\,f,g}$,}
\author[\,]{Mairi Sakellariadou$^{\,f}$,}
\author[\,]{Sotiris Sanidas$^{\,h}$,} 
\author[\,]{Lara Sousa$^{\,i,j}$,}
\author[\,]{Dani\`ele A.~Steer$^{\,a}$,}
\author[\,]{Jeremy M.~Wachter$^{\,c}$,}
\author[\,]{Sachiko Kuroyanagi$^{\,k}$}
\author[]{\\ \texttt{~~~~~~~~~~~~~~~~For the LISA Cosmology Working Group}\\}

\affiliation[a\,]{Laboratoire Astroparticule et Cosmologie, Universit\'e de Paris, 10 rue Alice Domon et Léonie Duquet, 75013 Paris, France}
\affiliation[b\,]{IKERBASQUE, Basque Foundation for Science, 48011, Bilbao, Spain}
\affiliation[c\,]{Department of Theoretical Physics, UPV/EHU, 48080, Bilbao, Spain}
\affiliation[d\,]{Institute of Physics, Laboratory of Particle Physics and Cosmology (LPPC), \'Ecole Polytechnique F\'ed\'erale de Lausanne (EPFL), CH-1015 Lausanne, Switzerland}
\affiliation[e\,]{Instituto de F\'isica Corpuscular (IFIC), CSIC-Universitat de Valencia, Spain}
\affiliation[f\,]{Theoretical Particle Physics and Cosmology Group, Physics Department, King’s College London, University of London, Strand, London WC2R 2LS, UK}
\affiliation[g\,]{Faculty of Physics, University of Warsaw ul.\ Pasteura 5, 02-093 Warsaw, Poland}
\affiliation[h\,]{Jodrell Bank Centre for Astrophysics, University of Manchester, Manchester, M13 9PL, United Kingdom}
\affiliation[i\,]{Instituto de Astrof\'{\i}sica e Ci\^encias do Espa{\c c}o, Universidade do Porto, CAUP, Rua das Estrelas, PT4150-762 Porto, Portugal}
\affiliation[j\,]{Centro de Astrof\'{\i}sica da Universidade do Porto, Rua das Estrelas, PT4150-762 Porto, Portugal}
\affiliation[k\,]{Department of Physics, Nagoya University, Chikusa, Nagoya 464-8602, Japan}

\abstract{
Cosmic string networks offer one of the best prospects for detection of cosmological gravitational waves (GWs). The combined incoherent GW emission of a large number of string loops leads to a stochastic GW background (SGWB), which encodes the properties of the string network. In this paper we analyze the ability of the Laser Interferometer Space Antenna (LISA) to measure this background, considering leading models of the string networks. We find that LISA will be able to probe cosmic strings with tensions $G\mu \gtrsim \mathcal{O}(10^{-17})$, improving by about $6$ orders of magnitude current pulsar timing arrays (PTA) constraints, and potentially $3$ orders of magnitude with respect to expected constraints from next generation PTA observatories. We include in our analysis possible modifications of the SGWB spectrum due to different hypotheses regarding cosmic history and the underlying physics of the string network. These include possible modifications in the SGWB spectrum due to changes in the number of relativistic degrees of freedom in the early Universe, the presence of a non-standard equation of state before the onset of radiation domination, or changes to the network dynamics due to a string inter-commutation probability less than unity. In the event of a detection, LISA's frequency band is well-positioned to probe such cosmic events. Our results constitute a thorough exploration of the cosmic string science that will be accessible to LISA.}

\begin{document}

\begin{figure}
\hskip11.cm 
\includegraphics[width = 0.29 \textwidth]{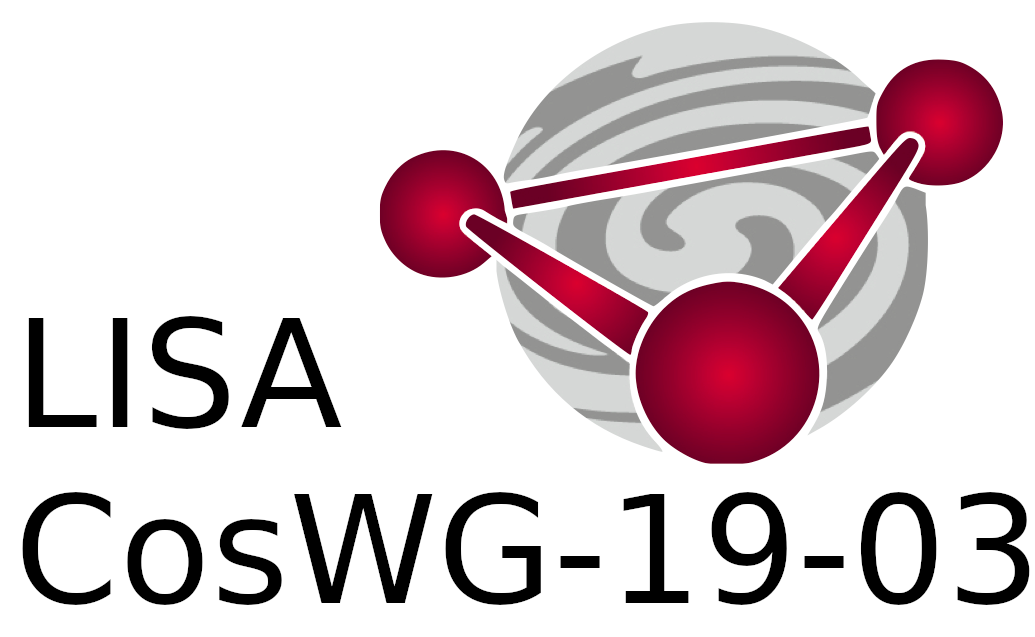}
\vspace*{0.5cm}
\end{figure}

\maketitle
\flushbottom

\section{Introduction}

The direct detection of gravitational waves (GWs) by the LIGO and Virgo network~\cite{Abbott:2016blz,Abbott:2016nmj,Abbott:2017vtc,Abbott:2017gyy,Abbott:2017oio} marks the dawn of a new era in astronomy, opening a unique window with which to observe the Universe. GWs carry invaluable information about the sources that created them --- which could be of astrophysical or cosmological origin --- since they propagate unimpeded through space. Gravitational waves therefore constitute one of the most promising new messengers with which we can probe aspects of the Universe so far undetermined by other means. 

One of the main targets of GW experiments is the detection of a stochastic gravitational wave background (SGWB) of cosmological origin. The most famous example of such a SGWB is the quasi--scale invariant background from inflation, due to quantum fluctuations~\cite{Grishchuk:1974ny,Starobinsky:1979ty, Rubakov:1982df,Fabbri:1983us}. This background is expected to be too small be to detectable by currently planned GW observatories. However, if axion-type species are present during inflation, potentially observable GWs can also be produced with a significant blue-tilt (see e.g.~\cite{Anber:2006xt,Sorbo:2011rz,Pajer:2013fsa,Adshead:2013qp,Adshead:2013nka,Maleknejad:2016qjz,Dimastrogiovanni:2016fuu,Namba:2015gja,Ferreira:2015omg,Peloso:2016gqs,Domcke:2016bkh,Caldwell:2017chz}, or~\cite{Bartolo:2016ami} for a general discussion on GWs from inflation). Furthermore, post-inflationary, early-universe phenomena can also generate GWs with a large amplitude, e.g.~a kination-dominated phase~\cite{Giovannini:1998bp,Giovannini:1999bh,Boyle:2007zx,Figueroa:2018twl,Figueroa:2019paj}, particle production during preheating~\cite{Easther:2006gt,GarciaBellido:2007dg,GarciaBellido:2007af,Dufaux:2007pt,Dufaux:2008dn,Dufaux:2010cf,Figueroa:2017vfa,Adshead:2018doq}, oscillon dynamics~\cite{Zhou:2013tsa,Antusch:2016con,Antusch:2017vga,Liu:2017hua,Amin:2018xfe}, strong first order phase transitions~\cite{Kamionkowski:1993fg,Caprini:2007xq,Huber:2008hg,Hindmarsh:2013xza,Hindmarsh:2015qta,Caprini:2015zlo,Cutting:2018tjt}, or cosmic defect networks~\cite{Vachaspati:1984gt,Sakellariadou:1990ne,Damour:2000wa,Damour:2001bk,Damour:2004kw,Figueroa:2012kw,Blanco-Pillado:2017oxo}. For a comprehensive review of SGWB signals of cosmological origin, see~\cite{Caprini:2018mtu}. In this paper, we focus on precisely one such cosmological source: cosmic strings. We investigate, in particular, the ability of the {\it Laser Interferometer Space Antenna} (LISA)~\cite{Audley:2017drz} --- which will be the first GW observatory in space --- to probe the SGWB emitted by a network of cosmic strings.

Cosmic strings are stable topological defect solutions of field theories~\cite{Nielsen:1973cs} which may have formed in symmetry breaking phase transitions in the early Universe~\cite{Kibble:1976sj,Jeannerot:2003qv}. Alternatively, they can be cosmologically stretched fundamental strings of String Theory, formed for instance at the end of brane inflation~\cite{Dvali:2003zj,Copeland:2003bj}. The energy per unit length of a string $\mu$, is of order $\eta^2$, where $\eta$ is a characteristic energy scale (for topological strings, the energy scale of the phase transition). In the simplest cases, the string tension is also of order $\mu$, and strings are relativistic objects that typically move at a considerable fraction of the speed of light. The combination of a high energy scale and a relativistic speed clearly indicates that strings should be considered a natural source of GWs.

A network of strings formed in the early Universe emits GWs throughout the history of the Universe, generating a SGWB from the superposition of many uncorrelated sources. In this paper, we forecast the constraints that LISA may put on the dimensionless combination $G\mu$ (where $G = 1/M_{p}^2$ is Newton's constant, and $M_p = 1.22\times 10^{19}$ GeV the Planck mass), which is related to the energy scale $\eta$ through
\be
G\mu \sim10^{-6} \left(\frac{\eta}{\rm{10^{16} ~GeV}}\right)^2,
\ee
and which parametrizes the gravitational interactions of the
string.

There are other ways one can hope to detect the presence of cosmic strings in the Universe that do not directly involve the observation of the GWs they generate. In fact, several potential observational signatures of cosmic string networks have been discussed in the literature, including anisotropies in Cosmic Microwave Background (CMB)~\cite{Ade:2013xla,Charnock:2016nzm,Lizarraga:2016onn,Ringeval:2010ca}, lensing events \cite{Vilenkin:1984ea,Bloomfield:2013jka}, and cosmic rays from the decay of strings into particle radiation \cite{Brandenberger:1986vj,Srednicki:1986xg,Bhattacharjee:1991zm,Damour:1996pv,Wichoski:1998kh,Peloso:2002rx,Sabancilar:2009sq,Vachaspati:2009kq,Long:2014mxa} (see \cite{Hindmarsh:1994re,Vilenkin:2000jqa,Sakellariadou:2006qs,Vachaspati:2015cma} for a review). Currently, CMB data from the Planck Satellite~\cite{Ade:2013xla} imply $G\mu < 10^{-7}$ for Nambu-Goto, Abelian-Higgs, and semi-local strings. The most stringent bounds, however, come from searches for the SGWB, with pulsar timing arrays (PTA) constraining $G\mu$ for Nambu-Goto strings to be $G\mu \lesssim 10^{-11}$~\cite{Blanco-Pillado:2017rnf,Ringeval:2017eww}, and LIGO-Virgo observations constraining it to as low as $G\mu < 2\times10^{-14}$, depending on the string network model~\cite{LIGOScientific:2019vic,Abbott:2017mem}. In this paper we show that LISA will be sensitive to string tensions with $G\mu \gtrsim 10^{-17}$ for Nambu-Goto strings, improving current upper bounds by $\sim 10$ orders of magnitude relative to CMB constraints, by $\sim 6$ orders of magnitude relative to current PTA constraints, and even by $\sim$3--4 orders of magnitude relative to future constraints from next generation of PTA experiments.

Since the characteristic width $\delta \sim 1/\eta$ of a cosmic string is generally much smaller than the horizon, in this paper we mainly assume that strings can be described by the Nambu-Goto (NG) action, which is the leading-order approximation when the curvature scale of the strings is much larger than their thickness.  We refer to such string as NG strings. Furthermore, we mainly focus on string networks without junctions; comments on cosmic superstring networks with junctions will be made in section \ref{subsec:intercomP}. With the NG action, one can study the evolution of a string network, from formation until the present time. While the basic picture is simple --- a string network is stretched by the cosmological expansion, and the motion of strings leads to multiple interactions and collisions between them --- in practice, this is a complicated problem which has been studied in depth in the literature.  Perhaps the most important conclusion of these studies is that the  cosmic string network reaches an attractor {\it scaling solution} in which its energy density remains a fixed fraction of the background energy density. One often assumes that when strings collide, they always intercommute, i.e., that they always ``exchange partners'' and reconnect after a collision.\footnote{This corresponds to an intercommutation probability $\mathcal{P}=1$, which we mainly assume throughout this paper. We comment briefly on $\mathcal{P}<1$, characteristic of cosmic superstrings, in section \ref{ss:inter}.}  As a result, closed loops are formed when a string self-intersects or two curved strings collide. Loops smaller than the horizon decouple from the cosmological evolution and oscillate under their own tension, slowly decaying into GWs. Indeed, in flat spacetime, one can show that loops have oscillating trajectories which are periodic in time. The relativistic nature of these strings typically leads to the formation of {\it cusps}, namely points where the string momentarily moves at the speed of light \cite{Turok:1984cn}. Moreover, the intersections of strings will generate discontinuities on their tangent vector known as {\it kinks}. All loops will contain kinks --- either as a result of the intercommutation that produced them, or as historical remnants of past intersections. Cusps and kinks generate gravitational wave bursts~\cite{Damour:2000wa,Damour:2001bk}, and these play a significant role in the SGWB emitted by string networks. (One should note that a complementary strategy to the detection of the stochastic background is therefore to search for such individual transient signals, see~\cite{Aasi:2013vna,Abbott:2017mem}.)
 
Other than sub-horizon string loops, the network also contains long strings that that stretch across a Hubble volume. These are either infinite or in the form of super-horizon loops, and are also expected to emit GWs. However, the dominant contribution is generically that produced by the superposition of radiation from many sub-horizon loops along each line of sight. Studying this SGWB and the possibility of observing it with the LISA constellation~\cite{Audley:2017drz} is the main focus of this paper. We argue that, even though the next round of pulsar timing observations could improve the constraint on the cosmic string tension $G\mu$ in the near future, this will not continue for long. Future tightening of these constraints will necessarily come from GW detectors operating in an intermediate frequency band. This is precisely due to the fact that the GW background expected from strings at lower energy scales will peak at these intermediate frequencies, which are out of reach of PTA experiments. We therefore conclude that LISA is the ideal instrument with which to search for cosmic strings in the future or, at the very least, to further improve constraints on cosmic string scenarios.

The paper is organized as follows. In Section~\ref{sec:SGWBfromStrings}, we briefly review the basic methods and relevant formulae with which to calculate the energy density spectrum of the SGWB emitted by sub-horizon loops in an evolving network of cosmic strings. In Section~\ref{sec:string-modeling}, we present different approaches developed in the literature to determine the loop number density, which is a fundamental quantity in the determination the SGWB from any string network. In section~\ref{sec:gw-from-strings}, we review the emission of GWs by individual strings, in particular the so-called `loop power spectrum' and the GW waveforms from bursts. These different results are put together in Section~\ref{sec:SpectrumGW}, where we characterize the spectral shape of the SGWB from a cosmic string network. We discuss different (potentially observable) features that can be imprinted in the SGWB spectrum, such as the details of radiation-to-matter transition, the number of relativistic degrees of freedom active during expansion, and the equation of state in the early Universe. In Section~\ref{sec:ProbingGWwithLISA}, we analyze in detail the ability of LISA to measure the spectrum of the SGWB from a network of cosmic strings, and in particular we determine the parameter space that is compatible with a detection. Finally, in Section~\ref{sec:overview}, we present an overview of our results and state our conclusions.

\section{The calculation of the SGWB from Cosmic Strings}\label{sec:SGWBfromStrings}

Several studies in the literature have calculated the SGWB generated by an evolving cosmic string network (see, e.g.,~\cite{Vilenkin:1981bx,Hogan:1984is,Vachaspati:1984gt,Accetta:1988bg,Bennett:1990ry,Caldwell:1991jj,Siemens:2006yp,DePies:2007bm,Olmez:2010bi,Sanidas:2012ee,Sanidas:2012tf,Binetruy:2012ze,Kuroyanagi:2012wm, Kuroyanagi:2012jf,Sousa:2013aaa,Blanco-Pillado:2013qja,Sousa:2014gka,Blanco-Pillado:2017oxo,Blanco-Pillado:2017rnf,Ringeval:2017eww,Cui:2018rwi,Jenkins:2018lvb}). This is often quantified in terms of the fraction of the critical density in GWs per logarithmic interval of frequency,
\be\label{eqn:theone}
\mathrm{\Omega}_{\rm gw}(t_0,f) = \frac{8 \pi G}{3H_0^2 }~f ~ \frac{d\rho_{\rm gw}}{df} (t_0, f)\,,
\ee
where $H_0$ is the Hubble parameter, and $\frac{d\rho_{gw}}{df}(t_0, f)$ is the energy density in gravitational waves per unit frequency $f$, observed today (at $t=t_0$).  The basic idea is that, given a GW frequency today, one must add up the GW emission from all the loops throughout the entire history of the Universe that contribute to that frequency. To do so, two different and complementary approaches have been developed in the literature, and the aim of this section is to introduce both of them.
(These two approaches are also discussed in more detail in section~\ref{sec:gw-from-strings}.)

Before doing so, we introduce the basic ingredients common to the two approaches. The first is the {\it number density $\nsf(l,t)$} of non-self-intersecting, sub-horizon, cosmic string loops of invariant length $l$ at cosmic time $t$.  These are the loops which, through their oscillations, contribute to the SGWB. When the network is scaling --- as it is in the radiation and matter eras --- $\nsf(l,t)$ can be estimated through different numerical and analytical techniques (see section~\ref{sec:string-modeling}).  Scaling, however, cannot be maintained during the radiation-to-matter transition, but analytical estimates can nonetheless be extended to this regime.
 
The second ingredient is the  loop power spectrum, namely the {\it power $P_{\rm gw}(f,l)$ emitted in GWs of frequency $f$ by a cosmic string loop of length $l$}.  It is clear that individual loops of a given length $l$ will radiate in different ways according to their shape. Hence either one can assume an average (or typical) gravitational loop power spectrum $P_{\rm gw}(f,l)$ determined numerically from simulations; or one can focus on particular events on the strings (cusps and kinks) for which $P_{\rm gw}(f,l)$ can be determined analytically.  

\subsection{Method I}\label{ssec:method1}

Let us write the power $P_{\rm gw}(f,l)$ in units of $G\mu^2$ and $l$ as
\be
    P_{\rm gw} (f,l)= G \mu^2 l ~P(fl),
\ee
where we have introduced a function $P(fl)$ which in principle takes a different form for each individual loop of length $l$, depending on its configuration.
The first method to calculate $\mathrm{\Omega}_{\rm gw}(t_0,f)$ assumes the existence of an averaged function, $P(fl)$, computed from an ensemble of loops of length $l$ obtained from simulations. Then the energy density in GWs observed at a particular frequency $f$ today is obtained by adding the amount of energy produced at each moment of cosmic evolution for loops of all sizes. On taking into account the redshift of frequencies from the moment of emission until today, one finds
\be\label{eqn:energyI}
   \frac{d\rho_{\rm gw}}{df}(t_0,f) = G\mu^2\int_0^{t_0}{dt \left(\frac{a(t)}{a_0}\right)^3 \int_0^\infty{dl ~ l~ \nsf(l,t)~ P \left(\frac{a_0}{a(t)} f l\right)}}\,,
\ee
where $a(t)$ is the scale factor which takes the value $a_0$ today.
In order to compute $\mathrm{\Omega}_{\rm gw}(t_0,f) $ from Eqs.~(\ref{eqn:theone}) and (\ref{eqn:energyI}), one must specify the cosmological model, the number density of loops $\nsf(l,t)$, and an average power spectrum $P(fl)$. This approach has been followed in e.g.~\cite{Vilenkin:1981bx,Hogan:1984is,Vachaspati:1984gt,Accetta:1988bg,Bennett:1990ry,Caldwell:1991jj,Sanidas:2012ee,Sanidas:2012tf,Binetruy:2012ze,Sousa:2013aaa,Blanco-Pillado:2013qja,Sousa:2014gka,Blanco-Pillado:2017oxo,Blanco-Pillado:2017rnf,Cui:2018rwi}.

\subsection{Method II}\label{ssec:method2}

At high frequencies $fl\gg 1$, $P_{\rm gw}(f,l)$ can be estimated analytically. Indeed, whatever the shape of the loop, one can show that the gravitational waveform sourced by a loop is dominated at high frequency by cusps, kinks, and kink-kink collisions. (See appendix~\ref{app:NG} for an overview of the Nambu-Goto equations and the precise definitions of cusps and kinks).  The form of $P_{\rm gw}(f,l)$ for these 3 types of events is discussed in section~\ref{sec:gw-from-strings}.

Cusps, kinks, and kink-kink collisions emit short bursts of GWs.  The contribution to the SGWB from the superposition of the unresolved signals from these three types of events is given by
\be\label{eqn:energyII}
\frac{d\rho_{\rm gw}}{df}(t_0,f) = f^2\int_{0}^{\infty} dz\int_{0}^\infty dl\,h^2(f,z,l)\,\frac{d^2R(z,l)}{dz dl}\,,
\ee
where $z$ is the redshift, $h(f,z,l)$ is the amplitude of the Fourier transform of the trace of the metric perturbation generated by each event, and $\frac{d^2R(z,l)}{dz dl}$ denotes the event rate per unit loop length and per unit redshift. This rate is directly proportional to $\nsf(l,t)$, and therefore one must know the number density of loops. This approach has been considered in refs.~\cite{Siemens:2006yp,Olmez:2010bi,Kuroyanagi:2012wm, Kuroyanagi:2012jf,Ringeval:2017eww,Abbott:2017mem,Jenkins:2018lvb}.

\subsection{Cosmology} \label{ssec:cosmology}

Finally, one must provide the details of the expansion history of the Universe. Unless specified otherwise, we assume a standard flat $\Lambda$CDM model. The Hubble rate reads
\be \label{eq:Hubblerate1}
H(z) = H_0 {\cal H}(z),
\ee
where
\be \label{eq:Hubblerate2}
{\cal H}(z) = \sqrt{\mathrm{\Omega}_{\Lambda} + \mathrm{\Omega}_{\rm mat} (1+z)^3 + \mathrm{\Omega}_{\rm rad} {\cal{G}}(z)(1+z)^4 }\,,
\ee
and we use Planck-2018 fiducial parameters~\cite{Aghanim:2018eyx},
\begin{align}
    H_0&=100 h ~{\rm  km/s/Mpc}\,, \nonumber \\
    h&=0.678\,,\nonumber \\
    \mathrm{\Omega}_{\rm mat}&=0.308\,,\\
    \mathrm{\Omega}_{\rm rad}&=9.1476\times 10^{-5}\,,\nonumber \\
    \mathrm{\Omega}_{\rm DE} &= 1-\mathrm{\Omega}_{\rm mat}-\mathrm{\Omega}_{\rm rad}\,. \nonumber
\end{align}
The function ${\cal G}(z)$, which takes into account the effective number of degrees of freedom $g_*(z)$ and the effective number of entropic degrees of freedom $g_S(z)$, is given by~\cite{Binetruy:2012ze}
\be \label{eq:DOFinHubble}
{\cal G}(z) = \frac{g_*(z) g_{\rm S}^{4/3}(0)}{g_*(0) g_{\rm S}^{4/3}(z)}\,.
\ee
Unless explicitly stated otherwise, we use the Standard Model numbers of degrees of freedom as given by  \texttt{microMEGAS}~\cite{Belanger:2018mqt}. We also make use of the following functions to describe proper distance
\be \label{eq:phir}
\varphi_r (z) = \int_0^z\frac{dz'}{{\cal H}(z')} dz
\ee
and proper volume
\be \label{eq:phiv}
\varphi_v (z) = \frac{4\pi \varphi_r^2(z)}{(1+z)^3{\cal H}(z) }\,.
\ee
We describe the imprint of the expansion history on the SGWB from cosmic string loops in section~\ref{sec:cosmological-imprint}. There we also discuss the effect of possible departures from this picture, including the impact of increasing the effective number of degrees of freedom in the early Universe as well as the impact of an equation-of-state different than that during radiation domination. 

The following sections describe in detail the different ingredients which enter into the calculation of the spectrum of gravitational waves.

\section{String network modeling}\label{sec:string-modeling}

We have mentioned earlier that one of the most important aspects of a cosmic string network is its ability to reach a scaling solution. Analytical modelling as well as early cosmic strings simulations demonstrated the approach of the long string network to this attractor regime~\cite{Bennett:1987vf,Albrecht:1989mk,Allen:1990tv}.  Loops, however, reach scaling over a longer time scale and therefore larger simulations are need to attain this regime. It is only more recently that Nambu-Goto simulations performed by two independent groups~\cite{Ringeval:2005kr,BlancoPillado:2011dq} have shown the existence of a population of scaling loops.

As outlined in section \ref{sec:SGWBfromStrings}, in order to calculate the spectrum of GWs expected today, a crucial input is the loop number density $\nsf(l,t)$ at all times $t$, since GWs are generated throughout the history of the cosmic string network. In order to extrapolate results from simulations, which run only over a finite time interval, to any moment in the history of the network, the scaling of loops is crucial since it implies that
\be
    \nsf(l,t) = t^{-4} \nsf(x)\,,
    \label{scn}
\ee
where $x = l/t$ is the ratio of the size of the loop to roughly the horizon scale. 

In order to obtain $\nsf(l,t)$, one approach is to determine the loop production function $f(l,t)~dl$, namely the number density of non-self-intersecting loops of lengths between $l$ and $l+dl$ produced per unit time, per unit volume,  which in scaling satisfies
\be
f(l,t)=t^{-5} f(x)\,.
\label{scf}
\ee
The number density of non-self-intersecting loops is then obtained by solving the Boltzmann equation for loops: loops are diluted with the expansion of the Universe, lose energy through GWs, and are sourced by loops being chopped off the infinite string network as described by $f(l,t)~dl$.\footnote{In principle loops could also collide with each other (to create larger, possibly self-intersecting, loops), leading to a more involved Boltzmann equation, see \cite{Copeland:1998na}. Loops could also rejoin the infinite string network, see \cite{Bennett:1985qt,Bennett:1986zn}. However, in \cite{BlancoPillado:2011dq}, this effect was shown not to be significant for non-self-intersecting
loops, and we will neglect it here.
} 
The loop number density can thus be computed by integrating the loop production function
\be\label{eqn:lnd}
    \nsf(l,t)  = \int_{t_{i}}^{t}{dt' f(l',t') \left(\frac{a(t')}{a(t)}\right)^3}\,,
\ee
where the effect of the expansion is explicitly seen through the dependence of the scale factor $a(t)$, and $l'$ (which is given below) contains information on the evolution of the length of the loop due to its gravitational decay from the time of formation $t'$ to the observation time $t$. More explicitly, assuming that, on average, the total power emitted by a loop is given by $\mathrm{\Gamma G} \mu^2$, where $\mathrm{\Gamma}$ is a dimensionless constant  (independent of the size and shape of a loop), then
\be
    l = l' + \mathrm{\Gamma} G\mu\,(t'-t)\,.
    \label{size}
\ee
(Namely, a loop with length $l'$ at time $t'$ has length $l$ at time $t>t'$.)  
As we discuss in more detail in later sections, the value of $\mathrm{\Gamma}$ is given by the sum of the GW power radiated at all frequencies, and therefore generally one would expect it to depend on the shape of the loop.  However, following the estimates from simple loops \cite{Vachaspati:1984gt,Burden:1985md,Garfinkle:1987yw} as well as the results obtained from recent simulations \cite{Blanco-Pillado:2017oxo}, in this paper we take $\mathrm{\Gamma}=50$.

The scaling loop number density for a power law cosmology parametrized by $a(t)\sim t^{\nu}$ can be obtained by combining Eqs.~(\ref{scn})-(\ref{size}). Changing variables from $t'$ to $x'=l/t'$ one finds~\cite{Blanco-Pillado:2013qja}
\be\label{eqn:BOS-integral}
    \nsf(x) =  \left[\frac{1}{\left(x + \mathrm{\Gamma}G \mu \right)^{3(1-\nu)+1}}\right] {\int_x^\infty \left(x' + \mathrm{\Gamma} G \mu \right)^{3 (1-\nu)} f(x')\,dx'}\,,
\ee
which can be easily computed once $f(x)$ is given.\footnote{In Eq.~(\ref{eqn:lnd}) we have assumed that $t\gg t_i$, meaning that the contribution from the loop distribution at the initial time $t_i$ can be neglected, and this is also the reason for the infinite upper limit in (\ref{eqn:BOS-integral}). Note that the Boltzmann equation may not always allow a scaling solution (see the analysis of \cite{Auclair:2019zoz}, valid for all $t\geq t_i$).}

Finally, we now relate $f(l,t)$ --- the loop production function for non-self-intersecting loops --- to the long string network with energy density $\rho_\infty$.  If we assume that the production of loops is the dominant energy loss mechanism of the long string network, then \cite{Vilenkin:2000jqa}
\be\label{eqn:cc}
   \frac{d \rho_{\infty}}{dt} = -2H(1+\vv^2)\rho_{\infty} -\mu\int_0^{\infty}{lf(l,t)dl}\,,
\ee
where $H=\dot{a}/a$ is the Hubble parameter and $\vv = \sqrt{\langle v^2 \rangle}$ is the root-mean-squared (RMS) velocity of the long strings strings. The first term in this equation describes the dilution of the long string energy density in an expanding Universe, while the second, proportional to the loop production function, takes into account the energy lost into loops. Loop production is essential to achieve the linear scaling of long strings, see e.g.~\cite{Vilenkin:2000jqa}.

In remainder of this section, we review three expressions for the loop number density $\nsf(l,t)$ which have been proposed in the literature for Nambu-Goto strings. Then, in section \ref{ssec:AH-FT}, we discuss the case of Abelian-Higgs string networks.

\subsection{Model I: analytic approach}\label{sec:VOS}

In the case of NG strings, the first expression for $\nsf(l,t)$ we consider is based on an analytic approach, which was initially developed by Kibble in \cite{Kibble:1984hp} and later extended in \cite{Caldwell:1991jj,DePies:2007bm,Sanidas:2012ee,Sousa:2013aaa}.  Here, the basic idea is that the loops produced by the long string network are described by a single free parameter (essentially the size of loops at formation),  
while eq.~(\ref{eqn:cc}) is used to determine the normalisation of the loop production function. 

As a first step in the determination of $\nsf(l,t)$, it is therefore necessary to have an analytical handle on the evolution of the long string energy density  $\rho_\infty$, and hence also of the RMS velocity $\vv$ appearing in eq.~(\ref{eqn:cc}).  To do so, following \cite{Sousa:2013aaa}, we use the successful Velocity-dependent One-Scale (VOS) model \cite{Martins:1996jp,Martins:2000cs} since this not only describes the scaling evolution of the long string network, but also its non-scaling evolution through the radiation-matter transition.\footnote{Here we reformulate the results presented in the original papers, in an attempt to unify our notation across the present paper.}
%
%
%
In terms of the characteristic length $L\equiv (\mu/\rho_\infty)^{1/2}$ --- which measures the average distance between long strings --- and $\vv$, the VOS equations of motion are \cite{Martins:1996jp,Martins:2000cs}
\bea
\frac{d\vv}{dt} & = & \left(1-\vv^ 2\right)\left[\frac{k(\vv)}{L}-2H\vv\right]\,,\label{eqn:vosv}\\
\frac{dL}{dt} & = & \left(1+\vv^ 2\right)HL+\frac{\cc}{2}\vv\,, \label{eqn:vosL}
\eea
where the constant phenomological parameter $\cc$ quantifies the efficiency of the loop-chopping mechanism. Indeed, since eq.~(\ref{eqn:vosL}) is simply eq.~(\ref{eqn:cc}) rewritten in terms of $L$, it follows that
\be
\cc \vv \frac{\rho_{\infty}}{L}=\mu\int_0^{\infty}{lf(l,t)dl}\,.
\label{eqn:cVOS}
\ee 
In eq.~(\ref{eqn:vosv}), the function $k(\vv)$ phenomenologically
accounts for the effects of small-scale structure (namely,  multiple kinks) on long strings, and we use the ansatz proposed in~\cite{Martins:2000cs}
\be \label{eqn:curavture}
k(\vv)=\frac{2\sqrt{2}}{\pi}\left(1-\vv^2\right)\left(1+2\sqrt{2}\vv^3\right)\frac{1-8\vv^6}{1+8\vv^6}\,,
\ee
which reproduces the expected asymptotic behaviour of $k(\vv)$ both in the relativistic and non-relativistic limits. 
The linear scaling of the long string network in the radiation- and matter-dominated backgrounds follows directly from eqs.~(\ref{eqn:vosv})-(\ref{eqn:vosL}) since the particular solutions
\be\label{eqn:scaling}
\frac{L}{t}=\sqrt{\frac{k(\vv)(k(\vv)+\cc)}{4\nu(1-\nu)}} \equiv \xi_s \qquad\mbox{with}\qquad \vv=\sqrt{\left(\frac{k(\vv)}{k(\vv)+\cc}\right)\left(\frac{1-\nu}{\nu} \right)}  \equiv \vv_s \,,
\ee
where the subscript $s$ stands for `scaling', are attractor solutions of these equations for $a\propto t^{\nu}$ and $0<\nu<1$.   More generally, eqs.~(\ref{eqn:vosv})-(\ref{eqn:vosL}) can be solved throughout any cosmological era, including the radiation-to-matter and matter-to-dark-energy transitions, and hence one can trace the evolution of cosmic string networks in a realistic cosmological background~\cite{Sousa:2013aaa}.  We note that although the VOS model only treats small-scale structure phenomenologically through $k(\vv)$, eqs.~(\ref{eqn:vosv},\ref{eqn:vosL}) were shown to provide an accurate description of the long string network evolution in both Nambu-Goto~\cite{Martins:2003vd} and Abelian-Higgs~\cite{Moore:2001px} simulations.\footnote{Several other analytical models, using more than one length scale, have been developed in an attempt to provide a description of a cosmic string network including small scales~\cite{Quashnock:1990qy,Kibble:1990ym,Austin:1993rg,Martins:2014kda}. These models can describe, in particular, the effects of gravitational radiation and gravitational backreaction. They generally contain a larger number of phenomenological parameters, and clearly the one describing the strength of gravitational backreaction cannot be calibrated with simulations (since simulations do not include gravitational backreaction). For an impact of the effect of gravitational backreaction on cosmic string dynamics, see ref.~\cite{Copeland:1999gn}.}  In the NG case, taking $\cc=0.23\pm0.04$ fits both radiation and matter era simulations~\cite{Martins:2000cs}. (Note that $\cc$ is the only free parameter in the VOS model.)

The second step is to relate the loop production function to the long string network as described by the VOS model.  Let us define $\xi\equiv \frac{L(t)}{t}$ and, as before, $x \equiv \frac{l}{t}$. Then, in terms of these variables, it follows from eq.~(\ref{eqn:cc}) (or alternatively eqs.~(\ref{eqn:vosL}) and (\ref{eqn:cVOS}))
that the loop production function satisfies
\be\label{eqn:energyconservation}
    \int_0^\infty{ x f(x) ~dx} = \frac{2}{\xi^2}\left[ 1- \nu (1+\vv^2)\right] = \cc \frac{\vv}{\xi^3}\,.
\ee
We now make the following assumption, characteristic of this model I:  {\it throughout cosmic history, all loops are assumed to be created with a length $l$ that is a fixed fraction of the characteristic length of the long string network}, namely $l = \alpha_L L$, with $\alpha_L <1$. Thus
\be
f(x) = \tilde C\delta(x - \alpha_L \xi)\,,
\label{eq:ffirst}
\ee
where from eq.~(\ref{eqn:energyconservation})
\be
\tilde C =  \frac{\cc}{\alpha_L}\frac{\vv}{\xi^4}\,
\label{c1}
\ee
with $\vv$ and $\xi=L/t$ being the solutions of the VOS equations (\ref{eqn:vosv})-(\ref{eqn:vosL}).
In fact, for reasons we now explain, we will consider a slightly modified form of $f(x)$, see (\ref{eqn:VOSlpf}) below. Indeed, note that the value of $\tilde{C}$ given in (\ref{c1}) is in fact an upper limit, since eq.~(\ref{eqn:energyconservation}) does not capture the fact that some of the energy from the long string network goes into redshifting of the peculiar velocities of loops: we account for this by introducing a reduction of the energy of loops by a factor of $f_r\sim\sqrt{2}$~\cite{Vilenkin:2000jqa}. Furthermore, the assumption that all loops are created with exactly the same size is not expected to capture the true distribution of loop lengths at formation. The effect of relaxing this assumption was studied in ref.~\cite{Sanidas:2012ee}, where it was found that considering a distribution of lengths generally leads to a decrease of the amplitude of the SGWB. To account for this effect, we introduce a second factor, ${\cal F}$, which in ref.~\cite{Blanco-Pillado:2013qja} was estimated to be $\mathcal{O} (0.1)$ for Nambu-Goto strings. Taking these correction factors into account, we rewrite the loop production function in (\ref{eq:ffirst}) as
\be\label{eqn:VOSlpf}
f(x) = \left(\frac{\mathcal{F}}{f_r}\right) {\tilde C} \delta\left(x - \alpha_L\xi\right)\equiv A ~ \delta\left(x - \alpha_L\xi\right)\, .
\ee
We stress that this expression is valid throughout cosmic history, even when the cosmic string network is not in a linear scaling regime (in this case $x$ and $\tilde{C}$ will be time-dependent).\footnote{Note that in the radiation era, the choice $l=\alpha_L L$ is equivalent to assuming that $l=\alpha t$, with $\alpha  =\alpha_L\xi_{r}$ (where $\xi_{r}$ is given in eq.~(\ref{eqn:scaling}) with $\nu=1/2$). In \cite{BlancoPillado:2011dq} the length of the loops produced in radiation and matter era simulations is estimated to be, respectively, $l_r=0.1t\simeq 0.33L_r$ and $l_m=0.18t\simeq 0.35L_m$. These values are well-described by a single value of $\al$ (more so than by a single value of $l/t$).}  Note also that since the length of a loop decreases with time due to gravitational radiation, see (\ref{size}), the maximum size of loops in this model is $l/t=\alpha_L \xi$.

The third and final step is to substitute eq.~(\ref{eqn:VOSlpf}) into eq.~(\ref{eqn:lnd}) in order to obtain the loop number density $\nsf(l,t)$ for all times, including during the radiation-to-matter and matter-to-dark-energy transitions.  Note that this in general requires solving the VOS equations  (\ref{eqn:vosL}) and (\ref{eqn:cVOS}). However, deep in the {\it radiation era} ($\nu=1/2$), the long string network is scaling and described by the VOS solutions (\ref{eqn:scaling}), namely $\xi_r=0.271$ and $v_r=0.662$, hence it follows that the loop distribution is given by
\be\label{eqn:VOSrad}
\nsf_r(x)=\frac{A_r}{\alpha}\frac{\left(\alpha+\mathrm{\Gamma} G\mu \right)^{3/2}}{\left(x+\mathrm{\Gamma} G\mu \right)^{5/2}}\,,
\ee 
with $A_r= 0.54$ (we fix ${\cal F}=0.1$, $f_r=\sqrt{2}$), and where we have defined $\alpha = \alpha_L \xi_r$. As noted above, this expression is only valid for $x\leq \alpha$. 
In a matter-only universe ($\nu=2/3$), the VOS scaling solutions (\ref{eqn:scaling}) give $\xi_m=0.625$ and $v_m=0.583$ and the loop distribution is
\be\label{eqn:VOSmat}
\nsf_m(x)=\frac{A_m}{\alpha_m}\frac{\alpha_m+\mathrm{\Gamma} G\mu}{\left(x+\mathrm{\Gamma}G\mu\right)^2}\,,
\ee
where $\alpha_m=\alpha_L \xi_m$, $A_m=0.039$ and $x\leq \alpha_m$. 

In section~\ref{ssec:rmtrans}, we use this analytical approach 
to estimate the effect  of the radiation-to-matter transition on the GW spectrum in the LISA frequency band.  In order to ease comparison with other loop distributions --- to which we now turn --- our results will be expressed in terms of $\alpha = \alpha_L \xi_r$ (and not the more natural parameter of this model, namely $\alpha_L$).  Furthermore, we also explore the effect of changing the loop size at formation, through $\alpha$, in section \ref{sssec:agnostic_loop_size}.

\subsection{Model II: simulation-inferred model of Blanco-Pillado, Olum, Shlaer (BOS)\label{ssec:BOS}}

The second loop number density distribution $\nsf(l,t)$ we consider was discussed in refs.~\cite{BlancoPillado:2011dq,Blanco-Pillado:2013qja}. There the authors performed NG simulations of cosmic string networks in the radiation and matter eras, and obtained the loop production functions for non-self-intersecting loops directly from these simulations. We now review these results and present the corresponding loop number density distributions in different cosmological eras.

\subsubsection{Radiation era}

In the radiation era, the results of ref.~\cite{Blanco-Pillado:2013qja} together with eq.~(\ref{eqn:BOS-integral}), lead to the following scaling number density of loops
\be\label{eqn:BOSrad}
\nsf_{r}(x) =  \frac{0.18}{\left(x + \mathrm{\Gamma} G \mu \right)^{5/2}}\,,
\ee
with a cutoff at the maximum size of a loop, $x\equiv l/t = 0.1$.  It then follows from (\ref{scn}) that the number density of loops in physical units reads
\be\label{eqn:BOSrad-physical}
\nsf_{r}(l,t) = \frac{0.18}{t^{3/2}\left(l + \mathrm{\Gamma} G \mu t \right)^{5/2}}\,,
\ee
with $l\leq 0.1 t$.  In ref.~\cite{Blanco-Pillado:2013qja}, the form of the loop production function was found numerically. It is not exactly a $\delta$-function, as assumed in Model I, however,  in \cite{Blanco-Pillado:2013qja} the precise form of the loop production function was argued not to be important, since for any function that respects the equation of energy balance given by eq.~(\ref{eqn:energyconservation}), the final form of the number density is universal. Hence one may argue that the most important piece of information from the simulation is the normalization factor of the loop number density in (\ref{eqn:BOSrad}). 

Comparing eq.~(\ref{eqn:BOSrad}) with eq.~(\ref{eqn:VOSrad}) shows the same power-law behaviour in the denominator, and furthermore fixing $\alpha=0.1$ (the maximum size of loops in these radiation-era numerical simulations), the normalization of eq.~(\ref{eqn:VOSrad})  yields 0.17 in the numerator, which is in good agreement with eq.~(\ref{eqn:BOSrad}). 

\subsubsection{Matter era}

The scaling distribution of loops from the radiation era survive past radiation-matter equality. The resulting number density of loops can be written in terms of the radiation density, $\mathrm{\Omega}_{\rm rad}$, and redshift $z$ as
\be\label{eqn:BOSradmatter}
\nsf_{\rm r,m}(l,t) = \frac{0.18\left(2\sqrt{\mathrm{\Omega}_{\rm rad}} \right)^{3/2}}{(l +\mathrm{\Gamma} G \mu t)^{5/2}}\left(1+z\right)^3\,,
\ee
where $t(z)$. This matches the previous expression (\ref{eqn:BOSrad-physical}) deep in the radiation era, and has the correct redshifting behaviour in the matter era.

Finally, loops are also produced once the network reaches scaling in the matter era. Following the results in ref.~\cite{Blanco-Pillado:2013qja}, the corresponding loop distribution is given by
\be\label{eqn:BOSmatter}
\nsf_{\rm m}(l,t) = \frac{0.27 - 0.45 (l/t)^{0.31}}{t^2 (l +\mathrm{\Gamma} G \mu t)^2}
\ee
for $l/t<0.18$. However, as we shall see in section \ref{sec:SpectrumGW}, for the values of the string tension $G\mu \lesssim 10^{-10}$ of relevance to LISA, the contribution of this population of loops to the SGWB is in fact negligible relative to (\ref{eqn:BOSrad-physical}) and (\ref{eqn:BOSradmatter}).

To summarise, in order to calculate the SGWB generated by cosmic string loops described by model II, eqs.~(\ref{eqn:BOSrad}-\ref{eqn:BOSmatter}) contain all the necessary information on the number density of loops at all times, from the formation of the cosmic string network until now.

\subsection{Model III: simulation-inferred model of Lorenz, Ringeval, Sakellariadou (LRS)}
\label{ss:LRS}

The final loop distribution we consider is that developed in \cite{Lorenz:2010sm} and based on a different NG string simulation to model II, namely~\cite{Ringeval:2005kr}. Furthermore, as opposed to ref.~\cite{Blanco-Pillado:2013qja}, the loop production function is not the quantity inferred from the simulation: rather, the authors~\cite{Ringeval:2005kr} extract directly the distribution of non-self-intersecting scaling loops from their simulation.  On scales  $x \gg \mathrm{\Gamma} G \mu$ they find\footnote{NG simulations do not include gravitational radiation, for which the characteristic scale is $\mathrm{\Gamma} G\mu$}
\be\label{eqn:LRSn}
\nsf(x) =  \frac{C_0}{x^{p}}\qquad\mathrm{for}\qquad x \gg \mathrm{\Gamma} G \mu\,,
\ee
where the values of the two constants $C_0$ and $p$ in the radiation and matter eras are
\bea
  \left. p = 2.60^{+0.21}_{-0.15}\right|_\mathrm{r} \,, &\qquad & \left. p = 2.41^{+0.08}_{-0.07} \right|_\mathrm{m}\,,\label{eqn:pLRS}\\
  \left. C_0 = 0.21^{-0.12}_{+0.13}\right|_\mathrm{r} \,, &\qquad & \left. C_0 = 0.09^{-0.03}_{+0.03} \right|_\mathrm{m}\,.\label{eqn:CLRS}
\eea
Compared with eq.~(\ref{eqn:BOSrad}), the radiation era solution has a similar amplitude but the power $p$ appears somewhat greater that $5/2$, with the indicated error bars. The power in the matter era differs than the one of Model II.

In order to extend the loop distribution (\ref{eqn:LRSn}) down to smaller scales, the authors of \cite{Lorenz:2010sm} solve the Boltzmann equation described in sec.~\ref{sec:string-modeling}, using a loop production function which itself is theoretically derived.  Indeed, following the analytical work of Polchinski, Rocha and collaborators~\cite{Polchinski:2006ee,Polchinski:2007rg,Dubath:2007mf}, it is modelled by a power law
$f(x)\propto x^{2\chi-3}$ for $x>x_{\rm c}$. Here
$x_{\rm c} \ll\mathrm{\Gamma} G\mu$ is a scale characteristic of gravitational backreaction, and was estimated in ref.~\cite{Polchinski:2007rg} to be given by
\be
x_c \equiv 20 (G\mu)^{1+2\chi}.
\label{eq:gammac}
\ee
The scaling loop distribution $\nsf(x)\;\forall x$ is then obtained\footnote{As shown in ref.~\cite{Lorenz:2010sm}, the form of the loop production function on smaller scales than $x_c$ is essentially unimportant to the final loop distribution.}  by substituting $f(x)\propto x^{2\chi-3}$ into equation (\ref{eqn:BOS-integral}), and finally the constant $\chi$ is fixed by comparing the resulting distribution on scales $x\gg \mathrm{\Gamma} G\mu$ to the numerically obtained distribution eq.~(\ref{eqn:LRSn}). One finds \cite{Lorenz:2010sm}\footnote{In ref.~\cite{Lorenz:2010sm}, it is assumed that $\chi<(3\nu-1)/2$; see ref.~\cite{Auclair:2019zoz} for an analysis in the case $\chi \geq (3\nu-1)/2$.}
\begin{equation}\label{eqn:chirsb}
  \chi_{\mathrm{r}} = 0.200^{+0.07}_{-0.10}\,,\qquad\chi_{\mathrm{m}}= 0.295^{+0.03}_{-0.04}\,.
\end{equation}
These values, together with Eq.~(\ref{eqn:CLRS}), fix all the parameters in the loop distribution $\forall x$.

The resulting distribution is given in ref.~\cite{Lorenz:2010sm}. In our analysis below, we have worked with the exact distribution given in that reference, but it is useful to present its approximate analytic form in the different regimes of loop length assuming scaling:
\begin{itemize}
    \item For loops with length scale large compared to $x_{\rm d}\equiv \mathrm{\Gamma} G\mu$:
        \begin{equation}\label{eqn:mod3_1}
            \nsf(x\gg x_{\rm d})\simeq \frac{C}{(x+x_{\rm d})^{3-2\chi}}\,,
        \end{equation}
    \item For loops with length scale smaller than $x_{\rm d}$, but larger than $x_{\rm c}$:
        \begin{equation}\label{eqn:mod3_2}
            \nsf(x_{ c}<x\ll x_{\rm d})\simeq \frac{C(3\nu-2\chi-1)}{2-2\chi}\frac{1}{x_{\rm d}}\frac{1}{x^{2(1-\chi)}}\,,
        \end{equation}
    \item For loops with length scale smaller than $x_c$, the distribution is flat:
        \begin{equation}\label{eqn:mod3_3}
            \nsf(x \ll x_{\rm c}\ll x_{\rm d})\simeq \frac{C(3\nu-2\chi-1)}{2-2\chi}\frac{1}{x_{\rm c}^{2(1-\chi)}}\frac{1}{x_{\rm d}}\,.
        \end{equation}
\end{itemize}
In the above,
\be
  C=C_0 (1-\nu)^{2-p}\,.
\ee

Relative to the BOS distribution, notice that the distribution in eq.~(\ref{eqn:mod3_3}) contains many more small loops (due to the inverse power of $x_c$ which is itself very small). In fact, these small loops dominate the stochastic GW spectrum at high frequencies, as was already discussed in ref.~\cite{Abbott:2017mem}, and hence can lead to very different constraints on $G\mu$ to that of the BOS model in the high frequency regime.  Indeed, the energy density in these small loops is very large, so the question of energy balance between the long string network and the loop distribution --- at least as described by eq.~(\ref{eqn:cc}) (with caveats mentioned in footnote 3) --- remains to be fully understood. 

\subsection{Abelian-Higgs field theory simulations}\label{ssec:AH-FT}

So far we have focused on Nambu-Goto strings which are infinitely thin. However, as mentioned in the introduction, cosmic strings are solitonic solutions of classical field theory models~\cite{Nielsen:1973cs} which means that, in principle, they can decay not only by releasing energy into gravitational waves but also directly into excitations of their elementary constituents.
 For this reason, a number of authors have simulated cosmic strings in different field theories. In this short section we review this work and the implications it may have for the loop-distribution $\nsf(l,t)$.

In ref.~\cite{Davis:1989nj} global (axionic) strings were studied and it was shown that decay into elementary constituents indeed takes place, in this case due to the coupling with the massless Goldstone mode present in the vacuum of the theory.
For local strings with no long-range interactions (and which, in the infinitely thin limit, are expected to be described by the NG action), the excitations in the vacuum are massive, and hence the expectation is that this radiation will be suppressed for long wavelength modes of the strings. This expectation is supported by simulations of individual oscillating strings~\cite{Martins:2003vd} and standing waves~\cite{Olum:1999sg}, which observe that massive particle radiation originates in high curvature regions of the string, e.g., in cusp-like regions where the string doubles back on itself~\cite{Olum:1998ag}. These simulations also support the fact that, except for the short bursts of energy, the strings evolve according to the Nambu-Goto equations of motion. Furthermore, recent simulations of individual loops in the Abelian-Higgs model ~\cite{Matsunami:2019fss} report that, for loops smaller than a critical length scale, the lifetimes of loops scale with the square of their lengths. Extrapolating their results to large loops, these authors conclude that for loops larger than the critical length scale, GW emission is expected to dominate over particle emission~\cite{Matsunami:2019fss}.

In contrast, large-scale field theory (FT) simulations of the whole network of strings~\cite{Vincent:1997cx,Hindmarsh:2008dw,Daverio:2015nva,Hindmarsh:2017qff} observe that the network of infinite strings reaches a scaling regime, thanks to energy loss into classical radiation of the scalar and gauge fields of the Abelian Higgs model. These large-scale simulations of cosmic string networks are therefore in disagreement with the above massive radiation arguments: they show the presence of extensive massive radiation being emitted, and so loops formed in these simulations decay within a Hubble time. This intriguing discrepancy has been under debate for the last $\sim$ 20 years, but the origin of this radiation is not currently understood.

The similarities and differences between FT and NG simulations of string networks can then be summarized as follows: the infinite strings are rather similar in curvature radius and length density, but loops decay into field modes in the FT simulations. In FT simulations the strings' energy density goes into massive modes of the fields, which are not part of the string network anymore. As a consequence, the string loops decay within a Hubble time, and hence do not contribute as a source of GWs all through cosmic history. In the Nambu-Goto picture, this channel does not exist, and instead the energy of the infinite strings goes into loops, which then decay into gravitational radiation. 

Our analysis in this paper is based on the NG classical evolution of strings. Hence we assume, as supported by NG simulations, that loops are formed throughout cosmic history, and they decay into GWs, as we describe in section~\ref{sec:gw-from-strings}. Our conclusions about the ability of LISA to measure a GW background from cosmic strings is therefore based on this fundamental assumption. 

\section{Gravitational wave emission from strings}\label{sec:gw-from-strings}

As outlined in section \ref{sec:SGWBfromStrings}, a crucial input into the calculation of the SGWB from cosmic strings is the loop power spectrum $P_{\rm gw}(f,l)$ (see Method I of section \ref{ssec:method1}).  Alternatively (for method II, section \ref{ssec:method2}), one requires both $h(f,z,l)$ and ${d^2R(z,l)}/{dz dl}$. 
Our purpose in this section is to determine these crucial quantities.  We also discuss the possibility of detecting individual burst events from loops, as well the contribution of long strings to the SGWB.

\subsection{GW loop power spectrum}
\label{sec:lps}

The power lost into gravitational radiation by an isolated loop of length $l$ can be calculated using the standard formulae in the weak gravity regime~\cite{Weinberg:1972kfs}. As a first approximation, we assume that the loop evolves in flat space, meaning that its evolution is periodic and radiation is only emitted at discrete frequencies, $\omega_n = 2\pi n/T$, where $T=l/2$ is the period of the loop, and $n=1,2,\ldots$.
Then the power emitted at frequency $\omega_n$ per solid angle is given by~\cite{Burden:1985md,Allen:1991bk}
\begin{equation}\label{eqn:dPndO}
    \frac{dP_n}{d\mathrm{\Omega}} = 8\pi G\mu^2n^2\left(|A_+|^2+|A_\times|^2\right)\,,
\end{equation}
where $A_{+,\times}$ are the amplitudes of the two gravitational wave polarizations. In a coordinate system in which ${\bf \hat{\mathrm{\Omega}}=\hat{z}}$, they are given by
\begin{subequations} \begin{align}
    A_+ = I^+_xI^-_x-I^+_yI^-_y\,,\\
    A_\times = I^+_xI^-_y+I^+_yI^-_x\,
\end{align} \end{subequations}
where the $I^\pm$'s are functions of the mode number $n$, and are related to the Fourier transform of the stress-energy tensor of the string. (The $\pm$ refer to the fact that the solutions of the NG equations in flat space are a superposition of left and right-moving waves, see appendix~\ref{app:NG}, where we also give the explicit expressions of $I^\pm$ in terms of these solutions.) These $I^\pm$ functions therefore encode the information about the geometric shape of the loop over its entire oscillation. Integration of eq.~(\ref{eqn:dPndO}) over the sphere around the loop yields the power, $P_n$, emitted in each mode for a particular loop.  If the loop contains cusps, kinks, and kink-kink collisions, then one can show generically that for large $n$, $P_n$ scales as $n^{-4/3}$, $n^{-5/3}$, and $n^{-2}$ respectively~\cite{Vachaspati:1984gt,Binetruy:2009vt}.   It is important to stress that the gravitational radiation from loops is quite anisotropic: for cusps, most of the radiation at high frequencies is localized within a small solid angle surrounding the cusp direction; for kinks, the radiation is emitted in a narrow strip on the celestial sphere around the loop (see section \ref{ssec:bursts} below).

The procedure outlined above has been used to calculate the power spectrum of certain simple analytic solutions of loops with a small number of harmonics~\cite{Vachaspati:1984gt,Burden:1985md,Garfinkle:1987yw}. The results are in general agreement with the analytic estimates from cusps and kinks given above. However, in order to calculate the stochastic gravitational wave spectrum from the whole network of loops, we need to estimate an {\it averaged loop power spectrum}, since different loops of different shapes (but same $l$) may have quite distinct power spectra.  One approach is to consider realistic loops obtained from a simulation. Furthermore, one could aim to go beyond the first approximation mentioned above (namely that the loop evolves in flat space), and consider how the shape of a loop changes due to the emission of gravitational radiation: that is, gravitational backreaction may be important to determine an accurate average power spectrum of loops. 

The effect of gravitational backreaction on the average loop power spectrum was first considered in ref.~\cite{Blanco-Pillado:2017oxo}. Starting from a representative group of $\sim 1000$ non-self-intersecting loops from a population of scaling loops in a large scale simulation, a simple toy model for backreaction was applied (the loops were smoothed at different scales), and finally the average power spectrum of the full family of loops was computed. The resulting spectrum --- which we denote as the {\it BOS spectrum} --- was found to be quite smooth, with a long tail well-described by $n^{-4/3}$, namely the high frequency region was dominated by cusps present on the smooth loops. Furthermore, the distribution of results for the total power, $\mathrm{\Gamma}= \sum_{n=1}^{\infty} P_n$, for those loops was found to be highly peaked at $\mathrm{\Gamma} \approx 50$.  
It is clear, however, that there is still some uncertainty in the accuracy of this power spectrum, since the smoothing procedure used in ref.~\cite{Blanco-Pillado:2015ana} only shares some of the key ingredients found in the results of recent studies of the gravitational backreaction~\cite{Wachter:2016rwc,Blanco-Pillado:2018ael,Chernoff:2018evo,Blanco-Pillado:2019nto}. These latter results indicate that some parts of the power spectrum could be affected differently by more realistic backreaction. 

In the following, we also consider the simple averaged loop power spectra that are determined exclusively from the frequency dependence of specific events (cusp, kinks and kink-kink collisions), namely
\begin{equation}\label{eqn:P-mono}
P_{ n}=\frac{\mathrm{\Gamma}}{\zeta(q)}n^{-q}\,,
\end{equation}
where $\zeta(q)$ is the Riemann zeta function, introduced as a normalization factor to enforce the total power of the loop to be equal to $\mathrm{\Gamma}=\sum_n P_n$. The parameter $q$ takes the values $4/3$,
$5/3$, or $2$ depending on whether the emission is dominated by cusps, kinks or kink-kink 
collision respectively.\footnote{This power spectrum should be understood as a discrete set of numbers that represent the power at each mode. We take this spectrum as it is, but we should bear in mind that this may not be a good approximation at low harmonics, where the structure of the entire loop becomes important.}  The sensitivity of the final SGWB to the value of $q$ will give us an indication of the robustness of our results relative to the uncertainty on $P_n$.


In terms of this average power spectrum $P_n$, Method I of section \ref{ssec:method1} yields the stochastic gravitational wave background as~\cite{Blanco-Pillado:2017oxo}
\begin{equation}\label{eqn:omega-method-1}
   \mathrm{\Omega}_{\rm gw}(\ln f) = \frac{8\pi G^2\mu^2 f}{3 H_0^2}\sum_{n=1}^\infty C_n(f) P_n\,,
\end{equation}
where 
\be\label{eqn:Cn}
C_n(f) = \frac{2n}{f^2 }\int_0^{\infty} \frac{dz}{H(z) (1+z)^6}~\nsf\left(\frac{2n} {(1+z)f},t(z)\right)\,,
\ee
which depends on the loop distribution through $\nsf\left(\frac{2n} {(1+z)f},t(z)\right)$, and on the assumed cosmological background through $H(z)$ and $t(z)$. As seen in section \ref{sec:string-modeling}, the number density of loops depends on the total power $\mathrm{\Gamma}$, and hence for consistency it is important to ensure that the average loop power spectrum is properly normalized.

\subsection{GW waveforms from bursts\label{ssec:bursts}}

As described in section~\ref{ssec:method2}, an alternative method to compute the SGWB from strings is to consider the
incoherent superposition of many bursts of from cusps, kinks and kink-kink collisions. The logarithmic Fourier transform of the corresponding waveforms from these individual events was calculated in ref.~\cite{Damour:2000wa,Damour:2001bk,Binetruy:2009vt}:
\be\label{eqn:hamplitude}
    h(l,z,f) = A_q(l,z) f^{-q} \,,
\ee
where
\be
    A_q(l,z) = g^{(q)}_1 \frac{G\mu H_0 l^{2-q}}{(1+z)^{q-1} \varphi_r(z)}\,.
\ee
Here $l$ is the length of the loop at redshift $z$ at which this particular event takes place, $\varphi_r(z)$ is a measure of the proper distance from the observer to the source (cf.~\eqref{eq:phir} in section~\ref{ssec:cosmology}), 
and as before $q=4/3,5/3$ and $2$, for cusps, kinks and kink-kink collisions, respectively. The numerical constant $g_1^{(q)}$ accounts for the fact that not all  cusps and kinks are identical (different cusps/kinks will have different geometry/sharpness), and this modulates the strength of the GW burst.  

As mentioned above, cusps and kinks radiate non-isotropically meaning that the above waveform is only valid for directions near the cusp or kink direction, and it should be cutoff on angles larger than~\cite{Damour:2001bk,Damour:2000wa}
\be
\theta_{\rm cutoff}(l,z,f) = \left(\frac{1}{g_2 f (1+z) l}\right)^{1/3}\,,
\ee
where $g_2=\frac{\sqrt{3}}{4}$. 
On taking into account the geometry of this beaming effect, the fraction $\Delta(l,z,f)$ of observable bursts  from cusps, kinks and kink-kink collisions is given by~\cite{Olmez:2010bi}
\be
\Delta(l,z,f) = \left(\frac{\theta_{\rm cutoff}(l,z,f)}{2}\right)^{3(2-q)} \Theta (1- \theta_{\rm cutoff}(l,z,f))~.
\ee
The rate of bursts, which is required for the calculation of the SGWB with method II (see eqn.(\ref{eqn:energyII})), is then given by~\cite{Olmez:2010bi}
\be\label{burstrate}
    \frac{d^2R(z,l)}{dz dl} = 2\varphi_v(z) ~ H_0^{-3} ~ \left(\frac{\nsf(l,t(z))}{l(1+z)}\right)~\Delta(l,z,f)\,,
\ee
where
 $\varphi_v(z)$ given in eq.~(\ref{eq:phiv}).

Finally we can collect these results together, and insert eqs.~(\ref{eqn:hamplitude}) and (\ref{burstrate}) into eq.~(\ref{eqn:energyII}) to find that the SGWB from Method II for a given type of burst to be given by
\begin{equation}\label{eqn:omega-method-2}
        \mathrm{\Omega}_{\rm gw}(\ln f)=\frac{\left(g_1^{(q)}\right)^2g_2^{-2+q}}{2^{5-3q}}\frac{2N_q(G\mu)^2(2\pi f)^3 }{3H_0^3}
        \int_0^\infty dx \int_{{z_{min}(x,f)}}^\infty dz \frac{ (ft(z))^{-2-q}}{(1+z)^{4+q}} x^{1-q} \frac{H_0}{H(z)}~\nsf(x)\,,
\end{equation}
where $N_q$ is the average number of bursts per oscillation in a loop, and $z_{min}(x,f)$ is the solution to $\theta_{\rm cutoff}(l,z_{min},f)=1$. 

Determining the average number of cusps and kinks for the loop network is a very non-trivial task and the subject of ongoing work, and given this uncertainty it is common to take $N_{c}=N_{k}=\mathcal{O}(1)$.
However, one can also consider a situation in which there will be contributions from all these types of events~\cite{Ringeval:2017eww,Abbott:2017mem,Jenkins:2018lvb}, namely $N_c$ number of cusps, $N_k$ kinks and $N_{kk}$ number of kink-kink collisions (with $N_{kk}=N_{k}^2/4$, on assuming that there are equal numbers of left- and right-going kinks). We then impose that the sum of all these events to the averaged total power of the loop, $\mathrm{\Gamma}$, is equal to the value used in the expression for the loop number density\footnote{This is basically the same type of condition used to impose the normalization of the loop power spectrum in eq. (\ref{eqn:P-mono}).}.  The resulting constraint between the set of parameters $g_1^{(q)},g_2,N_c,N_k$ and $\mathrm{\Gamma}$ is given in appendix~\ref{app:NG}.

Before presenting the results of the SGWB for the three loop distributions of section \ref{sec:SpectrumGW}, we finish this section by commenting on two important issues: the separation of strong infrequent bursts from the SGWB; and the potential contribution (which we have not discussed until now) of GWs being emitted by the long string network.

\subsection{Strong infrequent bursts}\label{ssec:popcorn}

The superposition of GW bursts from many cusps and/or kinks, as calculated in section \ref{ssec:bursts}, leads to a Gaussian stochastic background of GWs~\cite{Caprini:2018mtu}. However, strong infrequent bursts observed with a time interval greater than the period of GWs $\sim 1/f$ ($\sim 10^2 - 10^3{\rm s}$ for LISA) exhibit a non-Gaussian discontinuous signal, often referred to as ``popcorn-like''~\cite{Regimbau:2011bm}. Typically, these are bursts from low redshift, $z\ll 1$. If a burst occurs in our neighborhood and the amplitude is strong enough, then the signal can be identified individually by the burst detection pipeline. 

The non-Gaussian background from infrequent bursts is typically expected to be above the Gaussian background when strings have large tension and small initial loop size (e.g., $G\mu\sim 10^{-6}$ and $\alpha \sim 10^{-11}$ for LISA~\cite{Regimbau:2011bm}). Infrequent bursts are negligible for strings satisfying the current pulsar timing limit $G\mu < 10^{-11}$. We should therefore supplement, in principle, the expression for the SGWB calculation with a correction that suppresses the contribution from infrequent bursts. However, in practice, we have found that for large initial loop sizes, removing the rare burst has practically no effect on the present-day SGWB spectrum (see also ~\cite{Siemens:2006yp,Olmez:2010bi,Binetruy:2012ze,Blanco-Pillado:2017oxo}), at least when the number of cusps and kinks per loop oscillation period is $\mathcal{O}(1)$.

An interesting possibility is that the number of infrequent strong bursts could be greatly enhanced if we consider clustering of loops inside the dark-matter halo of our galaxy. This would mean that the loop number density could be enhanced by several orders of magnitude at the Sun's position greatly improving the detectability of single-burst events in LISA for $G\mu < 10^{-11}$~\cite{Chernoff:2017fll}.
It has also been shown recently using numerical relativity simulations that, for certain configurations, very small loops can emit GW bursts by collapsing to form black holes~\cite{Helfer:2018qgv}.
These are interesting proposals, but we do not discuss them further here, as they go beyond the scope of this paper, where we focus on the SGWB from a string network.

\subsection{Gravitational wave emission from long strings}\label{ssec:InfiniteStrings}

So far we have exclusively focused on the GW signal emitted from sub-horizon string loops. However, long strings (infinite and super-horizon loops) also emit GWs. One contribution to this signal is characterized by GWs emitted around the horizon scale at each time $t$, sourced by the anisotropic stress of the network~\cite{Krauss:1991qu,JonesSmith:2007ne,Fenu:2009qf,Figueroa:2012kw}. This background is actually expected to be emitted by any network of cosmic defects in scaling, independently of the topology and origin of the defects~\cite{Figueroa:2012kw}, and hence represents an irreducible background generated by any type of viable defect network that has reached scaling. However, in the case of NG cosmic string networks, this background represents a sub-dominant signal compared to the GW background emitted from the loops. In the case of field-theory strings (for which simulations to date indicate the absence of ``stable'' loops), it is instead the only GW signal (and hence the dominant one) emitted by the network.

The GW energy density spectrum of this secondary background produced by long strings is predicted to be exactly scale-invariant for the modes emitted during radiation domination~\cite{Figueroa:2012kw}. At the level of the power spectrum, this background mimics therefore the spectral shape of the dominant signal from the loop decay (see discussion in section~\ref{subsec:BasicShape}), except with a much smaller amplitude. Even though the shape of the power spectrum of this irreducible GW background is well understood theoretically, its ultimate amplitude depends on the fine details of the so called \textit{unequal-time-correlator} of the network's energy-momentum tensor. Unfortunately, this correlator can only be obtained accurately from sufficiently fine lattice simulations of defect networks. It is therefore difficult to assess at this point whether this background can be detectable with LISA. In the case of global defects, the scale-invariant GW power spectrum has been estimated in ref.~\cite{JonesSmith:2007ne,Fenu:2009qf}. The amplitude of the spectral plateau has been calibrated in lattice field theory simulations for global strings\footnote{In the more interesting case of Abelian-Higgs lattice field theory simulations, there is unfortunately no quantification of the amplitude of this background.} as~\cite{Figueroa:2012kw}
\begin{equation}\label{eq:GWSOSFplateau}
    h^2\mathrm{\Omega}_{\rm gw}\simeq 4\times 10^4\, h^2\mathrm{\Omega}_{\rm rad}(G\mu)^2\,.
\end{equation}
Even though the numerical prefactor is much larger than unity, the quadratic scaling proportional to $(G\mu)^2$ suppresses significantly this background (see e.g.~\cite{Buchmuller:2013lra} for a comparison of this amplitude and that of the GW signal emitted from the decay of Nambu-Goto string loops during RD). This amplitude is clearly subdominant when compared to the amplitude of the dominant GW signal from the loops, which scales as $(G\mu)^{1/2}$ (see eq.~(\ref{eqn:plateau-model1}) and the discussion in section~\ref{sssec:HughFreqPlateau}). A proper assessment of the ability of LISA to detect the power spectrum of this stochastic background requires further results not available yet; namely, lattice simulations of cosmic networks with a larger dynamical range.

One can also consider the contribution to the GW spectrum coming from the accumulation of small-scale structure on long strings. These kinks are the product of the multiple intercommutations that infinite strings suffer over the course of their cosmological evolution, and were noticed early on in numerical simulation of cosmic networks~\cite{Bennett:1987vf,Sakellariadou:1990nd}. The emission of GW from individual infinite strings modulated by kinks has been calculated in refs.~\cite{Sakellariadou:1990ne,Hindmarsh:1990xi}. Using these results, one can also compute the spectrum produced by these kinks on a network assuming the simple model in which their characteristic scale is given by $\alpha t$. At high frequencies one can then estimate that the radiation-era plateau of this contribution should be~\cite{Vilenkin:2000jqa}
\be
    h^2\mathrm{\Omega}_{\rm gw} \simeq  \frac{128 \pi^2}{3 \xi^2 \alpha}  \, h^2\mathrm{\Omega}_{\rm rad}(G\mu)^2\,,
\ee
which for $\alpha\approx 0.1$ and $\xi_{\rm r}=0.271$ shows a rough agreement with the value obtained from field theory simulations. On the other hand, recently, ref.~\cite{Matsui:2019obe} has calculated the GW spectrum produced by kink-kink collisions on long strings, and found that the amplitude is larger than in previous estimates. This is because the characteristic scale $\alpha$ turns out to be much smaller than $0.1$ according to their semi-analytic estimation of the kink number distribution. 

As all these backgrounds are clearly sub-dominant against the SGWB from loops, we will not consider them in the following analysis of the paper (except for a brief discussion of the bispectrum in section~\ref{ssec:BisectrumInfiniteStrings}).

\section{Spectrum of the SGWB from cosmic string loops}\label{sec:SpectrumGW}

As discussed above, a string network evolves towards a scaling solution in which its energy density is simply proportional to the total background energy density $ \mathrm{\Omega}_{\infty}\propto G\mu~\mathrm{\Omega}_{\rm tot}$. The string network constantly produces loops which then emit GWs, and follow the background expansion instead of simply redshifting (which would correspond to $\mathrm{\Omega}_{\infty}\propto a^{-2}$ for static infinite strings). This continuous emission and tracking with expansion makes cosmic strings a perfect source for probing the expansion history of the Universe. In fact, in this section we show that all features visible in a stochastic GW frequency spectrum can be traced to a corresponding characteristic period in the evolution of the Universe.

We start our analysis by determining the basic shape of the SGWB spectrum over many decades in frequency, once a given loop number density distribution is chosen. We then study the impact of loops created relatively recently, that is, during the radiation-to-matter transition. Finally, we characterize the impact of extra degrees of freedom and other possible modifications of the equation of state in the very early Universe.

\subsection{Basic spectral shape}\label{subsec:BasicShape}

\begin{figure}
    \centering
    \includegraphics[height=9cm]{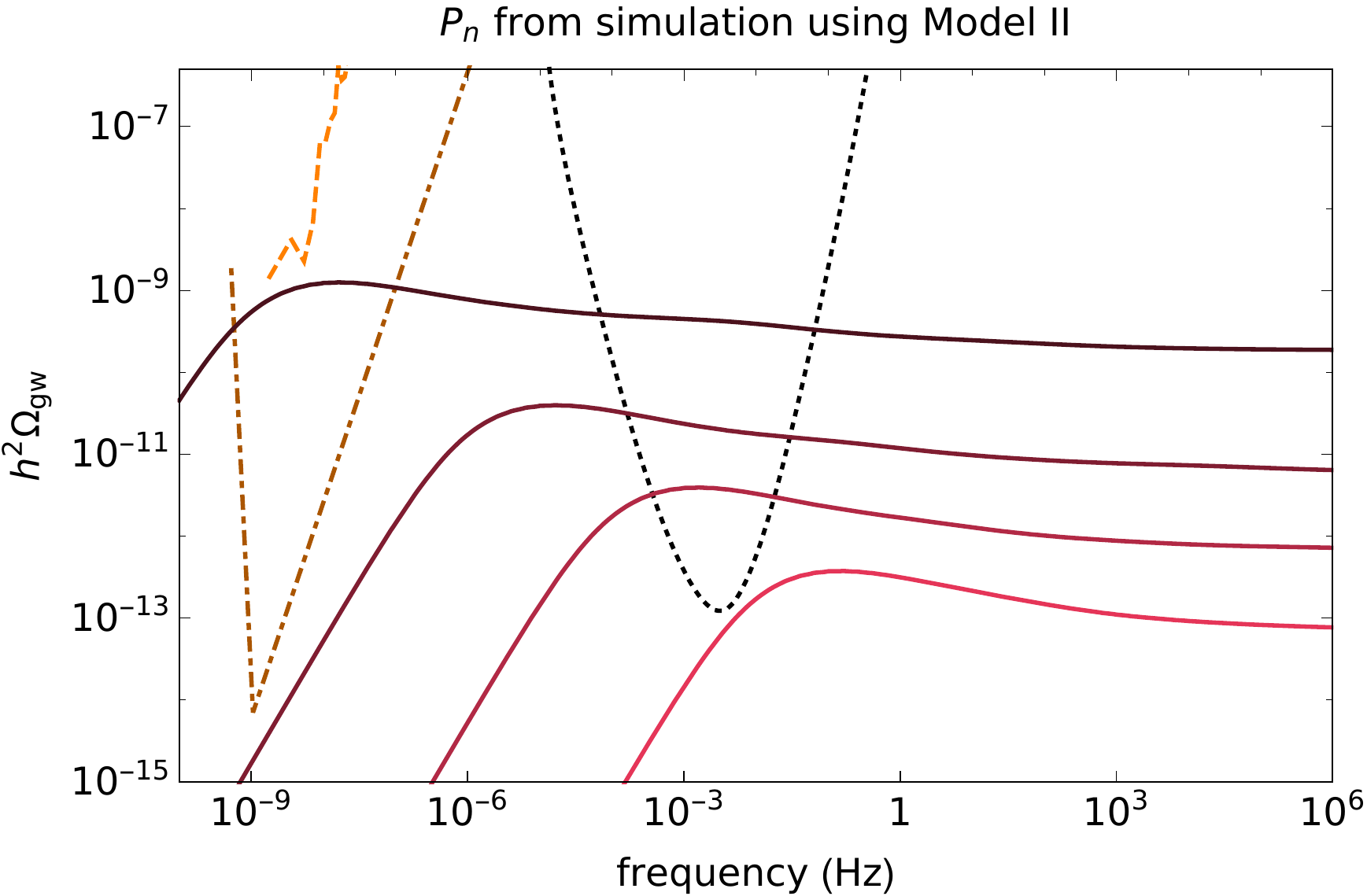}
    \caption{Cosmic string SGWB curves (all in red) near various relevant values of $G\mu$. The dashed orange curve is the EPTA sensitivity, and the darkest red curve just below is for $G\mu=10^{-10}$. The dash-dotted dark orange curve is the (projected) SKA sensitivity, and the dark red curve just below is for $G\mu=10^{-13}$. The dotted black curve is the LISA PLS; the red curve whose peak passes through it, and the light red curve just below, are for $G\mu=10^{-15}$ and $10^{-17}$ respectively. The $P_n$ are inferred from simulation~\cite{Blanco-Pillado:2017oxo}, and the loop number density is from Model II.}\label{fig:fiducialSGWB}
\end{figure}

\begin{figure}
    \centering
    \includegraphics[height=9cm]{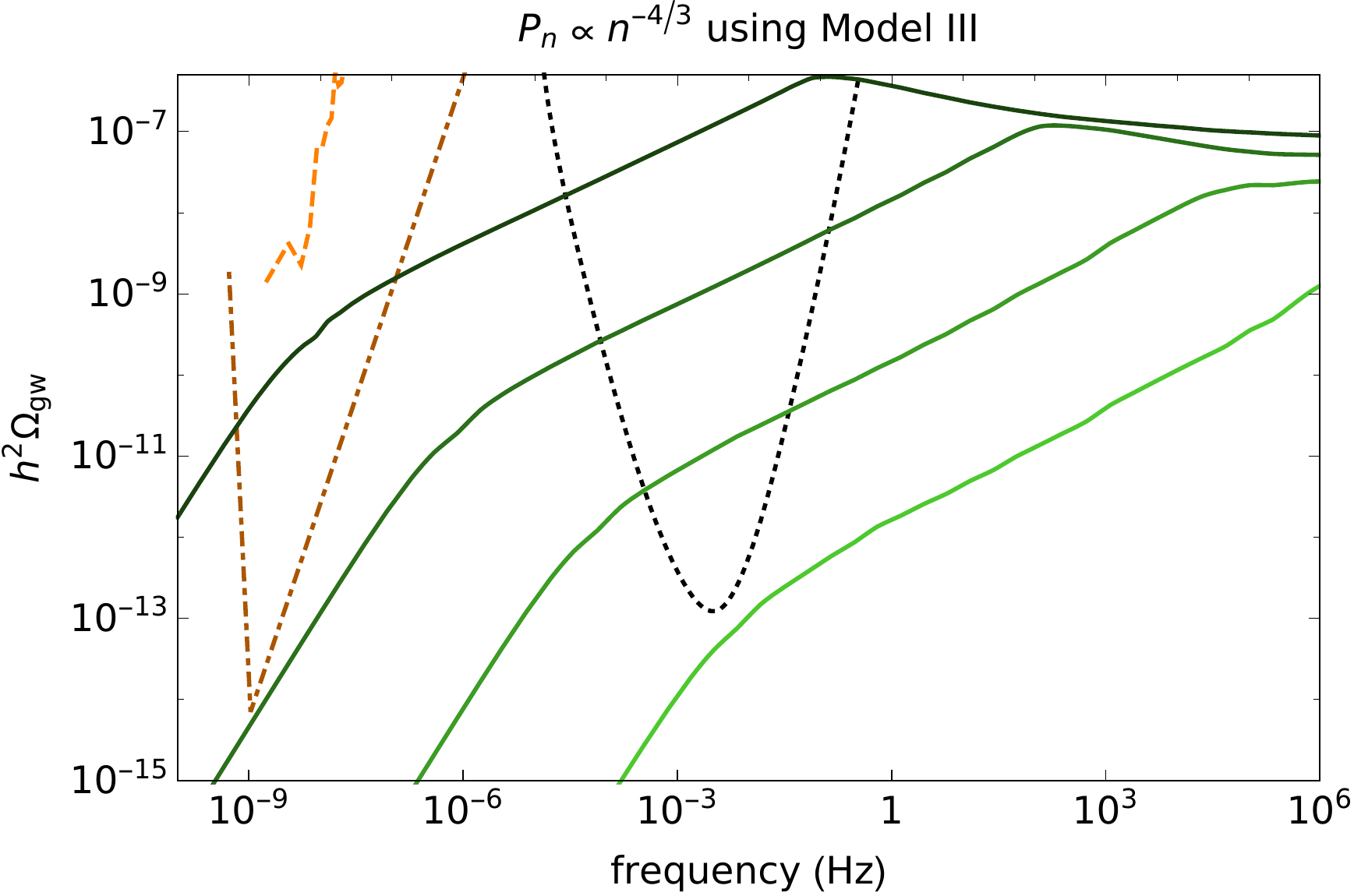}
     \caption{Idem as figure~\ref{fig:fiducialSGWB}, but with $P_n\propto n^{-4/3}$ and using the loop number density from Model III~\cite{Lorenz:2010sm}.  }
   \label{fig:combined-LRS-large}
   \end{figure}

The expressions given in eqs.~(\ref{eqn:omega-method-1}) and (\ref{eqn:omega-method-2}) can be used to compute the SGWB. In the following, we set $\Gamma = 50$ and we use Method I (eq.~(\ref{eqn:omega-method-1})) to generate several SGWB spectra for different values of $G\mu$. To illustrate our main points, we first take the loop number density from Model II and the loop power spectrum denoted by BOS in section \ref{sec:lps}: the results are shown in figure~\ref{fig:fiducialSGWB}. In figure~\ref{fig:combined-LRS-large} on the other hand, we use the loop number density from Model III and a monochromatic spectrum of cusps only ($q=4/3$). The difference between
these results for the same value of $G\mu$ comes almost
entirely from the different loop number density of small loops
in these models, as discussed in section \ref{ss:LRS}.

These figures plot the SGWB for a set of representative values of $G\mu$ together with the current sensitivity curves for EPTA pulsar timing collaboration~\cite{Kramer:2013kea}, as well as the projected curves for the SKA~\cite{Janssen:2014dka} and LISA~\cite{Audley:2017drz} collaborations. In particular, we show the spectrum of $G\mu = 10^{-10}$ as being the order of the bound on the string tension coming from current observations of pulsar timing arrays (PTAs). This bound should be improved in the next few years. However, as the limit on $ \mathrm{\Omega}_{\rm gw}$ becomes stronger and one probes lower values of the tension, one can see that the peak of the SGWB moves towards high frequencies and outside of the PTA frequency bands. This makes future bounds less strong than one would have thought because the PTA frequency band will then be at the steep section of the SGWB curve. Eventually, the SKA collaboration will become more competitive, potentially setting a bound of $G\mu=2\times 10^{-13}$, three orders of magnitude stronger than current PTA constraints.

An important point to make here is that if any of these observations detect a SGWB, one will probably have to wait for LISA before one can elucidate the origin of such background. It is therefore interesting to see that if $G\mu$ is in the range of values accessible by PTA experiments, the higher-frequency part of the SGWB signal will be well within LISA's sensitivity curve. The spectrum for $G\mu = 10^{-13}$ in figure~\ref{fig:fiducialSGWB} shows how such a curve might appear in LISA.


Looking at the curves for $G\mu = 10^{-15}$ and $10^{-17}$ in figure~\ref{fig:fiducialSGWB}, it is clear that for lower string tensions, PTA-type experiments become irrelevant for detecting a background and at this level LISA becomes the right instrument to probe these light strings~\cite{Blanco-Pillado:2017rnf}. The ``bump'' of the SGWB will pass directly through the LISA sensitivity band, as shown for $G\mu = 10^{-15}$, and $G\mu = 10^{-17}$ is the order of the lower bound on tension that LISA will set.

\subsubsection{The high frequency regime}\label{sssec:HughFreqPlateau}

As we can see from the SGWB curves shown in figure~\ref{fig:fiducialSGWB}, the spectrum becomes flat at very high frequencies. This can be understood analytically using a scaling number density of loops as well as a simplified cosmological background that describes the evolution of the Universe deep in the radiation era. The combination of these two facts allows us to find an expression (following Method I) for the spectrum of the form
\begin{equation}
    \mathrm{\Omega}^{\rm plateau}_\text{\rm gw}(\ln f)_I = \frac{64\pi G^2\mu^2 \mathrm{\Omega}_{\rm rad}}{3} \left(\sum_{n=1}^\infty P_n\right) \left( \int{dx ~\nsf_r(x) } \right)\,.
\end{equation}
This shows that indeed the SGWB is flat in this regime, but also that it only depends on two properties of the network of strings: the averaged total power emitted by a loop, and the total number of loops. Applying this to Model II, we find
\begin{equation}\label{eqn:plateau-model1}
    \mathrm{\Omega}^{\rm plateau}_{\rm gw}(\ln f)_I \approx 8.04\mathrm{\Omega}_{\rm rad}\sqrt{\frac{G\mu}{\mathrm{\Gamma}}}\,.
\end{equation}
This is a relevant result as it tells us that the value of the high-frequency plateau only depends on $G\mu$ and the total $\mathrm{\Gamma}$. In particular, it does not depend on the exact form of the loop's power spectrum, nor on if the GW emission is dominated by cusps or kinks, but rather depends only on the total radiation emitted by the loops.

Similarly, we can perform the same kind of computation using Method II. Starting with eq.~(\ref{eqn:omega-method-2}), and taking the cosmological background to be in the radiation era, we find \footnote{Note that in order to make this comparison, one should express the result in terms of the total power emitted ${\mathrm{\Gamma}}$. We give in appendix~\ref{app:NG} the calculation of $\mathrm{\Gamma}$ in terms of the parameters $N_q, g_1, g_2$.} a good agreement for the plateau with the expression found in eq.~(\ref{eqn:plateau-model1}). This is expected, given that the plateau only depends on quantities that must be identical in both methods. However, given the different nature of the calculations performed in both methods, this is a good consistency check.


\subsection{Radiation-to-matter transition}\label{ssec:rmtrans}

Numerical simulations studying the strings scaling have typically been performed in fixed backgrounds: pure radiation domination and pure matter domination~\cite{Blanco-Pillado:2013qja,Blanco-Pillado:2017oxo,Lorenz:2010sm,Ringeval:2017eww}. The usual simplified approach would be to just switch between the two loop distributions at radiation--matter equality; however, in reality we expect the network to smoothly evolve between the two regimes. In fact, the string network evolves rather slowly and, as pointed out in~\cite{Sousa:2013aaa}, does not reach scaling regime with matter background up until the current accelerated expansion starts. This may have a significant impact on the number density of loops in the matter era.

\begin{figure}
\centering
\includegraphics[height=7.5cm]{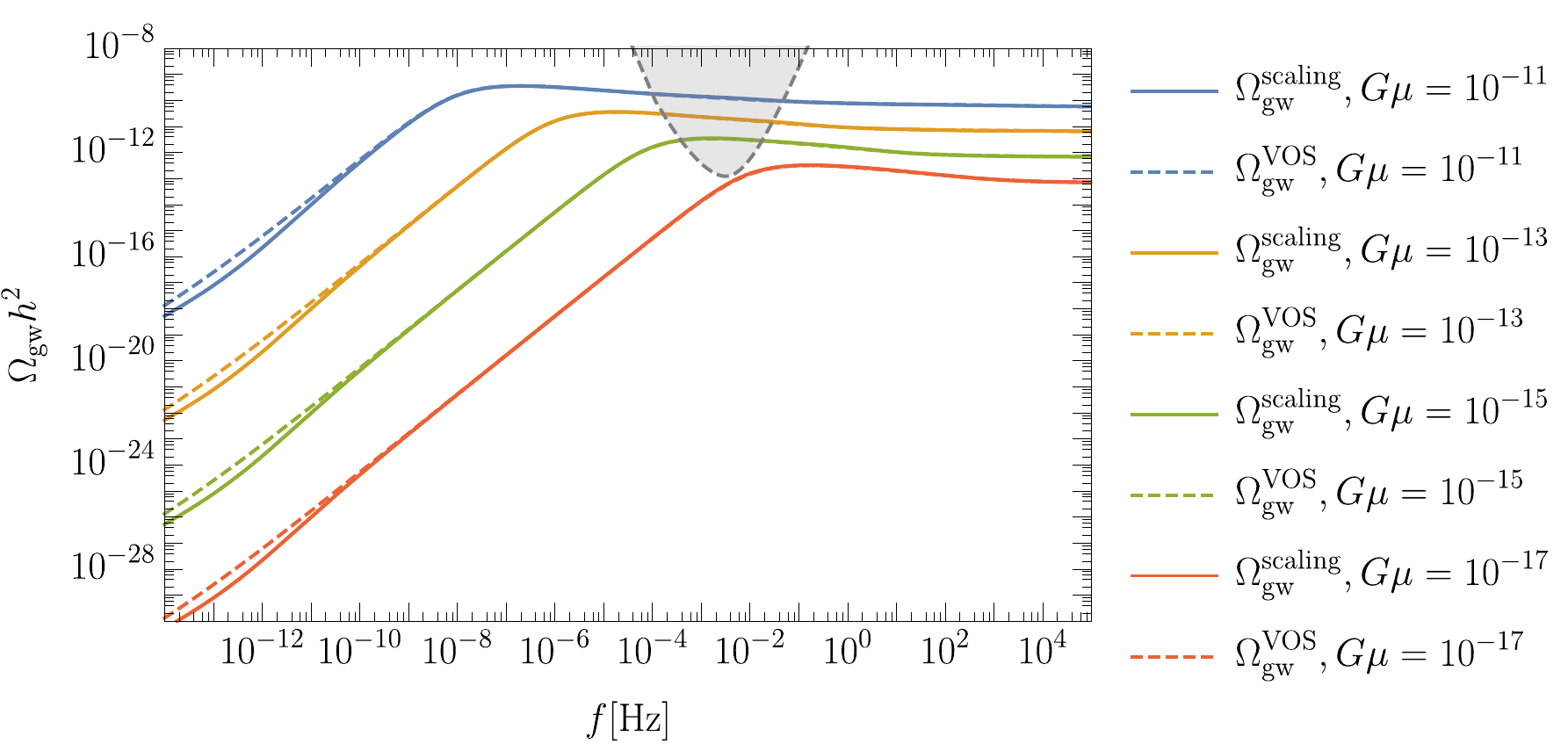}
\caption{Examples of spectra for several values of $G\mu$ and $\alpha=10^{-1}$ using both the full VOS solution (with {\it VOS} superscript) and assuming the network is always in scaling through eqs.~(\ref{eqn:VOSrad},\ref{eqn:VOSmat}) (with {\it scaling} superscript). The gray area indicates LISA sensitivity.}\label{fig:GWRDtoMD}
\end{figure}

We can study the impact of more careful modelling of the transition using the analytical model discussed in section~\ref{sec:VOS}. In figure \ref{fig:GWRDtoMD} we compare results coming from the full evolution of the loop density, eq.~(\ref{eqn:BOS-integral}), with the simplified spectrum obtained performing an instantaneous switch between the scaling results in matter domination (eq.~(\ref{eqn:VOSmat})), and radiation domination (eq.~(\ref{eqn:VOSrad})).
Figure~\ref{fig:GWRDtoMD} shows examples of spectra for several values of $G\mu$ and $\alpha=0.1$ using both prescriptions. As we can see, the inclusion of a smooth radiation-to-matter transition only modifies the spectrum significantly at very low frequencies $f\lesssim 10^{-10}~{\text{Hz}}$, outside of the LISA band. The reason is that it is only at these very low frequencies that the signal is dominated by loops created in the matter era~\cite{Blanco-Pillado:2017oxo}.

Even though the peak in the spectrum always appears due to matter domination, for low $G\mu\lesssim 10^{-11}$ it is only created by redshifting of GWs and the loop density in the matter background, while the loops dominantly contributing are formed much earlier, deep in the radiation era. With this we can safely conclude that for large loops $\alpha=0.1$ suggested by recent simulations, the modelling of the radiation-to-matter transition is irrelevant in the LISA sensitivity window. In section~\ref{sssec:agnostic_loop_size} we discuss how this situation may change if we assumed smaller loop sizes.

\subsection{Variation of the relativistic degrees of freedom}\label{sec:cosmological-imprint}

Another feature in the expansion rate of the Universe that would leave a clear signature in the stochastic GW spectrum of a cosmic string network is a modification in the number of relativistic degrees of freedom~\cite{Bennett:1985qt}. Whenever the temperature of the plasma forming our radiation background drops below the mass of a certain particle, that species will annihilate, injecting energy into the plasma and temporarily reducing its rate of cooling. This effect is automatically included in our calculation by solving the Friedman equation, eq.~(\ref{eq:Hubblerate2}), which includes the impact of changes in the number of degrees of freedom on the expansion rate through eq.~(\ref{eq:DOFinHubble}).

\begin{figure}
\centering
\includegraphics[height=7.5cm]{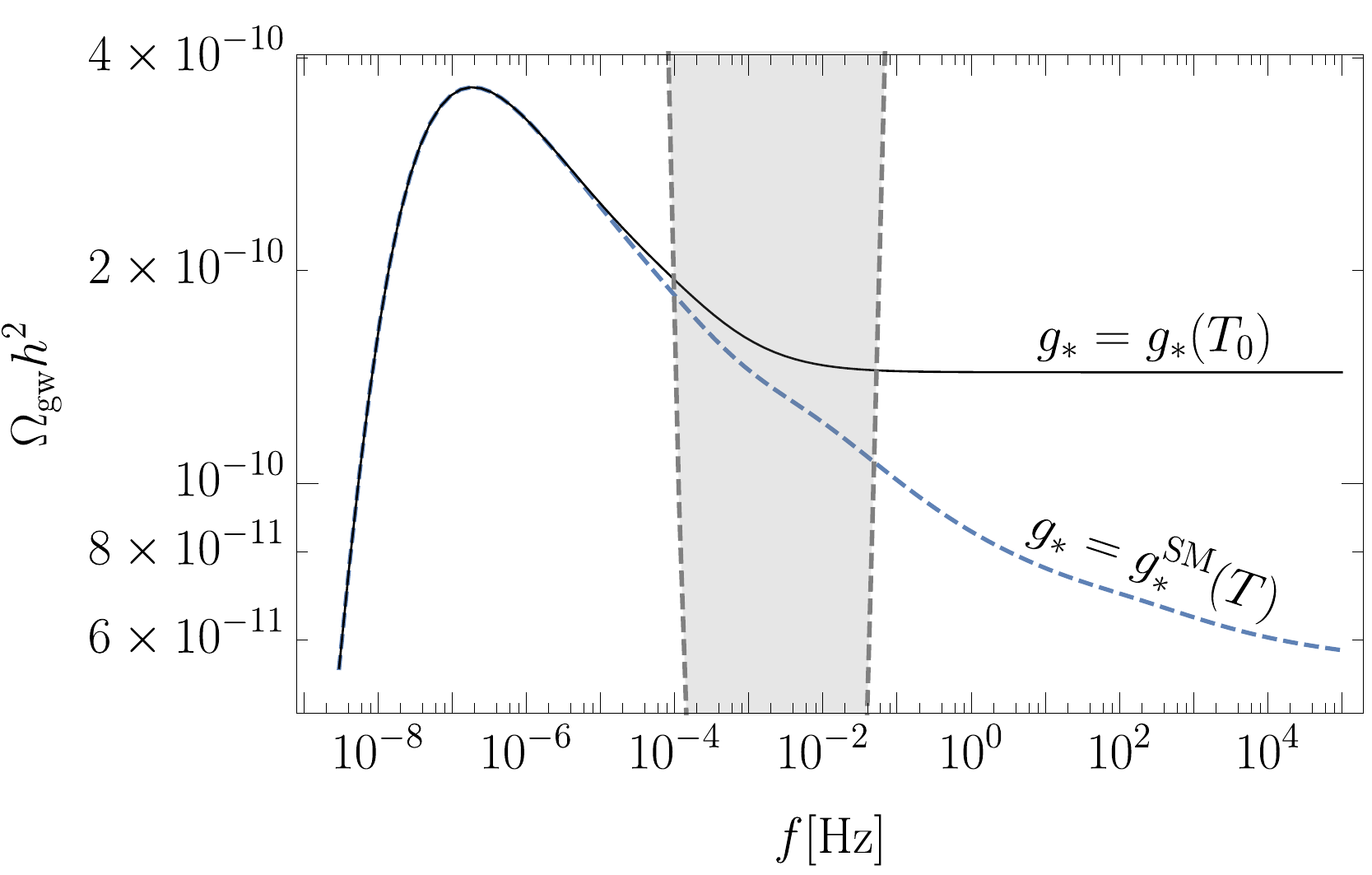}
\caption{Examples of spectra with $G\mu=10^{-11}$ assuming a constant number of degrees of freedom (black solid line) and standard cosmology with SM particle content (blue dashed line). The gray area indicates LISA sensitivity.}
\label{fig:dofexplot0}
\end{figure}

We show the impact of including this variation in figure~\ref{fig:dofexplot0}, which shows both the result obtained using the Standard Model number of degrees of freedom and just a constant value.
As we can see, the modification of number of degrees of freedom produces smooth variations in the spectrum at the frequency corresponding to the temperature of the modification. The most prominent of these variations in the spectrum correspond to electron-positron annihilation at $T\approx 200\,{\rm keV}$ where the lines first separate, the QCD phase transition at $f\approx 10^{-2} $ Hz ($T \approx 100\,{\rm GeV}$), and the electroweak scale at $f \gsim 10^2$ Hz. This means LISA could probe the QCD equation of state and other SM processes through their impact on the stochastic background from cosmic strings~\cite{Hajkarim:2019csy}.

Crucially, this effect would also potentially allow us to observe extra degrees of freedom (DOF) from beyond the standard model~\cite{Cui:2018rwi,Caldwell:2018giq,Blanco-Pillado:2017oxo,Battye:1997ji}. Had the number of DOF increased by a factor of $\Delta g_*$, that would have created another smooth step, changing the value of the plateau at the corresponding frequency by
\be\label{eqn:gstar_effect}
    \frac{\mathrm{\Omega}_{\rm gw}}{\mathrm{\Omega}_{\rm gw}^{\rm SM}} \approx
    \left(\frac{g_*^{\rm SM}}{g_*^{\rm SM}+\Delta g_*} \right)^{1/3}\,,
\ee
where $g_*^{\rm SM}$ and $g_{*S}^{\rm SM}$ are the number of degrees of freedom and the number of entropy degrees of freedom, both calculated in the standard model.

\begin{figure}
\centering
\includegraphics[height=7.5cm]{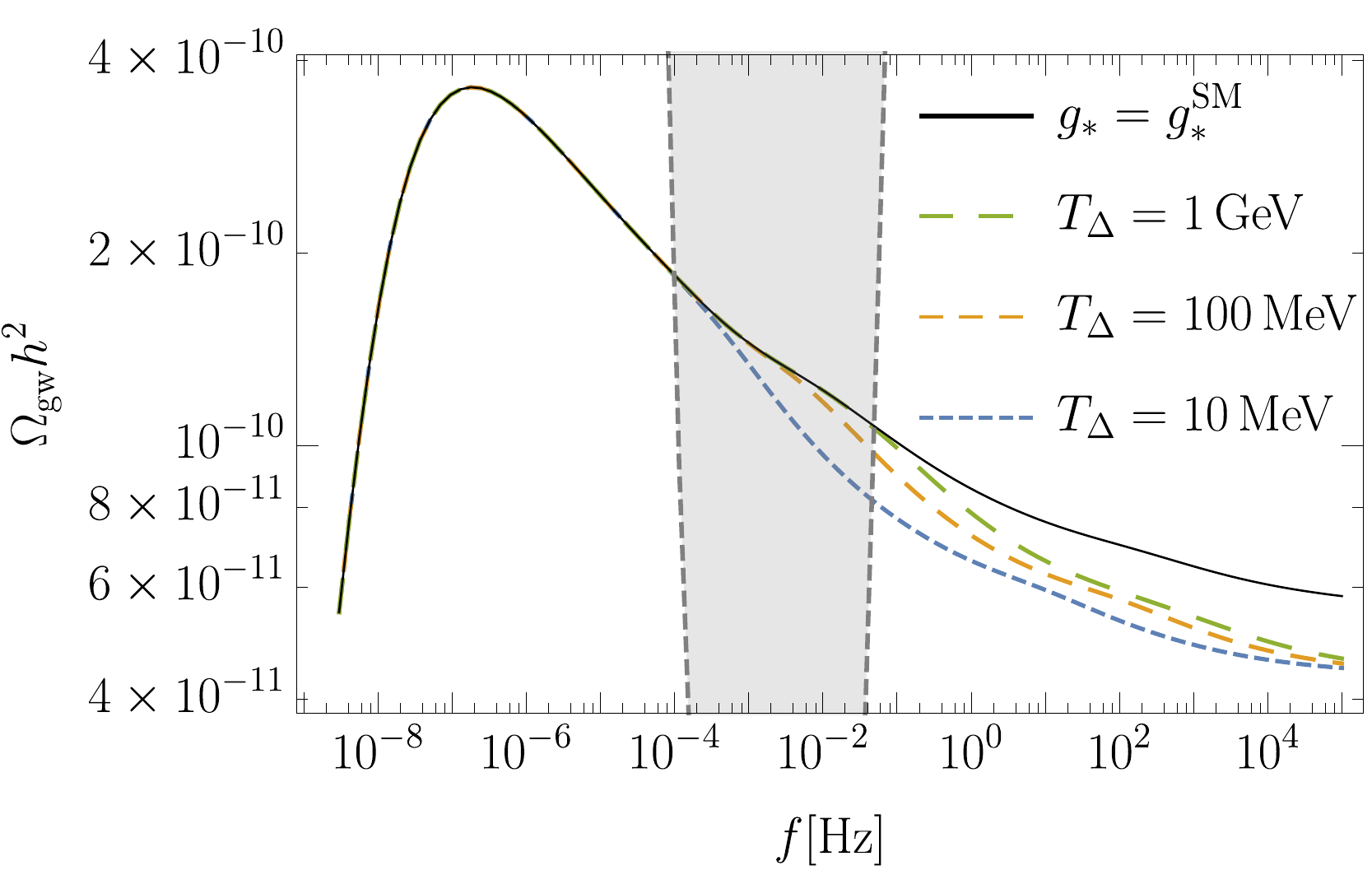}
\caption{Examples of spectra with $G\mu=10^{-11}$ in standard cosmology (black solid line) and several spectra in cosmological evolution with $\Delta g_*=100$ new degrees of freedom annihilating at at the range of temperatures of  interest for LISA. The gray area indicates LISA sensitivity.}\label{fig:dofexplot}
\end{figure}

We can numerically check that the frequency corresponding to a modification of the expansion rate occurring at a temperature $T_{\Delta}$ is given by~\cite{Cui:2018rwi}
\be \label{eqn:fdeltaforlargealpha}
f_{\Delta}=
  (8.67\times 10^{-3} \, {\rm Hz})\,
\left(\frac{T_\Delta}{\rm GeV} \right)
\left(\frac{0.1\times 50\times 10^{-11}}{\alpha\,\mathrm{\Gamma}\,G\mu}\right)^{1/2}
  \left(\frac{g_*^{\rm SM}(T_\Delta)}{g_*^{\rm SM}(T_0)}\right)^\frac{8}{6} \left(\frac{g_{*S}^{\rm SM}(T_0)}{g_{*S}^{\rm SM}(T_\Delta)}\right)^{-\frac{7}{6}}\,.
\ee
Using this estimate, we can see that LISA frequency band corresponds to probing temperatures of the order of a few GeV. It is important to point out this could lead to a significant improvement over the current probes of the expansion rate, which can reach only to the BBN temperature of a few MeV, which is still 3 orders of magnitude lower than the potential of a cosmic string signal at LISA. In figure~\ref{fig:dofexplot}, we show examples of a cosmic string stochastic background in standard cosmology with $G\mu=10^{-11}$ and several modifications with $\Delta g_*=100$ new degrees of freedom dropping out of equilibrium at the range of temperatures of interest for LISA.

\subsection{Probing the cosmological equation of state at early times}

The reasoning used in the last subsection also clearly applies to more dramatic modifications of cosmology in which the expansion at early times is dominated by something other than primordial radiation. A typical example here would be an early period of matter domination~\cite{Moroi:1999zb} after which the matter decays and the Universe resumes the standard radiation dominated expansion. Another example, so-called \textit{kination}~\cite{Joyce:1997fc,Giovannini:1998bp,Giovannini:1999bh,Salati:2002md,Chung:2007vz,Poulin:2018dzj,Figueroa:2018twl,Figueroa:2019paj}, is a period of domination of a new constituent of energy that redshifts faster than radiation and eventually becomes subdominant, avoiding any conflict with late time experiments.

\begin{figure}
\centering
\includegraphics[height=6.3cm]{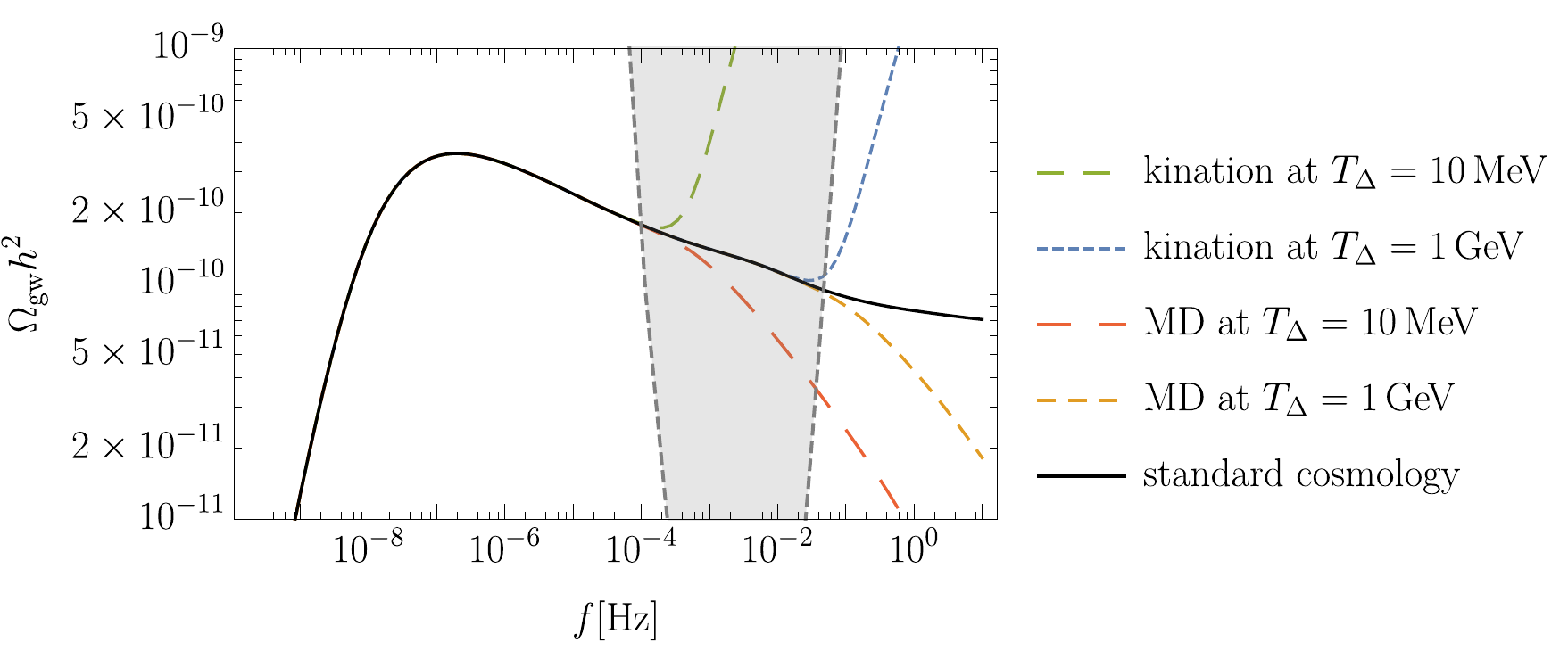}
\caption{Examples of spectra with $G\mu=10^{-11}$ in standard cosmology (black solid line) and several spectra in cosmological evolution with a period of early matter domination as well as kination ending in the range of temperatures of interest for LISA. The black dashed line indicates LISA sensitivity.}\label{fig:modexplot}
\end{figure}

Observation of the plateau of GW spectrum from a cosmic string network would indeed verify radiation domination up to $T_\Delta$ from eq.~(\ref{eqn:fdeltaforlargealpha}). If any nonstandard behaviour is observed, it can be traced back to the underlying modification. Simply expanding eq.~(\ref{eqn:omega-method-1}) at high frequencies, we can check that the impact of modified redshifting in a background $H^2\propto a^{-\beta}$ would simply lead to
\begin{equation}\label{eqn:analyticalspectrum}
    \mathrm{\Omega}_{\rm gw}(f>f_{\Delta})\propto
    \begin{cases}
        f^{\frac{8-2 \beta}{2- \beta}} \quad \beta \geq \frac{10}{3}, \\
        f^{-1} \quad \quad \beta<\frac{10}{3},
    \end{cases}
\end{equation}
behaviour above $T_{\Delta}$. An early period of matter domination corresponds to $\beta=3$. However, for expansion in the early Universe with any $\beta<\frac{10}{3}$, the emission from the string network is in fact subdominant to the tail of the distribution produced at later times. This leads to some degeneracy, and in fact if the network simply achieved scaling only at that time after their production~\cite{Vilenkin:2000jqa}, or if scaling was delayed due to the network having been diluted by inflation~\cite{Guedes:2018afo}, it would also amount to the lower case in eq.~(\ref{eqn:analyticalspectrum}). For scenarios with a new energy constituent redshifting faster than radiation (that is, $\beta>4$), the spectrum rises after $T_{\Delta}$, which generically gives better observational prospects.\footnote{Note, however, that the current Planck data puts a constraint on the total energy density of gravitational waves~\cite{Henrot-Versille:2014jua} $\int \mathrm{\Omega}_{\rm gw}\,h^2 d(\ln f)<3.8\times10^{-6}$. Consequently, any deviations from radiation domination with $\beta >4$ should have had a limited duration to avoid overproduction of GWs.} We show examples of the resulting spectra with the range of $T_{\Delta}$ of interest for LISA in various modified cosmologies in Fig~\ref{fig:modexplot}.

\section{Probing the SGWB from a string network with LISA}\label{sec:ProbingGWwithLISA}


The {\it Laser Interferometer Space Antenna} (LISA)~\cite{Audley:2017drz}, approved by the European Space Agency (ESA) in 2017, will be the first GW observatory in space. The final configuration adopted by the collaboration has been fixed to six links, 2.5 million km--length arms, and 4 years nominal duration, possibly extensible to 10 years. LISA will have the ability to search for GWs around the currently unexplored milli-Hertz regime.

To characterize the detectability of a SGWB with a spectrum described by a single power law (fully characterized by an amplitude and slope), ref.~\cite{Thrane:2013oya} introduced a very useful concept: the power law sensitivity curve (PLS) of a detector. This is a method that exploits the fact that the sensitivity of a detector increases when integrating a SGWB signal over frequency, in addition to integrating over time. The PLS curve is a spectral representation that  graphically quantifies, for a given signal-to-noise ratio, the ability of a detector to measure a SGWB with a power law (PL) spectrum. Searches by current GW experiments (by LIGO/VIRGO and PTAs) on power spectra of the form $\Omega_{\rm gw}(f) = A f^n$ have not succeeded in a detection, and hence they only provide upper bounds on the amplitude $A$ for different fixed values of the spectral index $n$~\cite{LIGOScientific:2019vic,Lentati:2015qwp,Arzoumanian:2018saf}. 

Recently, the LISA collaboration has presented a new technique for a systematic reconstruction of a SGWB signal without assuming a power-law spectrum~\cite{Caprini:2019pxz}. The idea is to first separate the entire LISA band into smaller frequency bins, and then to reconstruct a given arbitrary signal within each bin, where it can be well-approximated in terms of a power law. The method can reconstruct, in this way, signals with arbitrary spectral shapes, taking into due account instrumental noise at each frequency bin. Such analysis would be particularly appropriate for our case, as the spectral shape of the SGWB from cosmic string loops is not a simple plateau (and hence not a simple power law) for the lowest $G\mu$ values that LISA can probe. Furthermore, the spectrum can also exhibit scale-dependent features within the LISA frequency band, such as whenever there are changes in the number of relativistic degrees of freedom and/or the early Universe equation-of-state.

As this multi-band analysis technique has only very recently become available ($\sim 1-2$ weeks before the completion of this draft), in the present paper we will simply continue using as a criterion for detection that the spectrum of the SGWB from the string loops must be equal or above the PLS curve. We will use the LISA PLS as introduced by ref.~\cite{Thrane:2013oya}, but using the most updated LISA sensitivity curves based on the final configuration of LISA and new knowledge on its noise (see ref.~\cite{LISA_docs} for all relevant LISA documents up to date, and in particular ref.~\cite{LISA_PLS} for a direct download of the {\it Science Requirements Document}). The details of the updated LISA PLS curve used in this work can be found in Ref.~\cite{Caprini:2019pxz}. Whenever we claim detection of a given spectrum of the SGWB from cosmic string loops, if the spectrum is a power-law within the LISA sensitivity band, this can be interpreted as a detection of a SGWB after 3 years of collecting data (which corresponds to 4 years of LISA operation), with a signal-to-noise (S/N) ratio $\geq 10$. If the shape is more complicated than a simple power law, a more elaborated analysis following ref.~\cite{Caprini:2019pxz} is required to assess the S/N for a given detection, see also~\cite{Karnesis:2019mph}. In the present work, we simply quantify the parameter space compatible with a detection, but do not quantify the S/N associated to such detection, neither we reconstruct such parameter space with appropriate statistical techniques. We leave these aspects for future work.

\subsection{Projected constraints on the string tension}\label{ssec:gmu-constraints}

\begin{figure}
    \centering
    \includegraphics[height=8.cm]{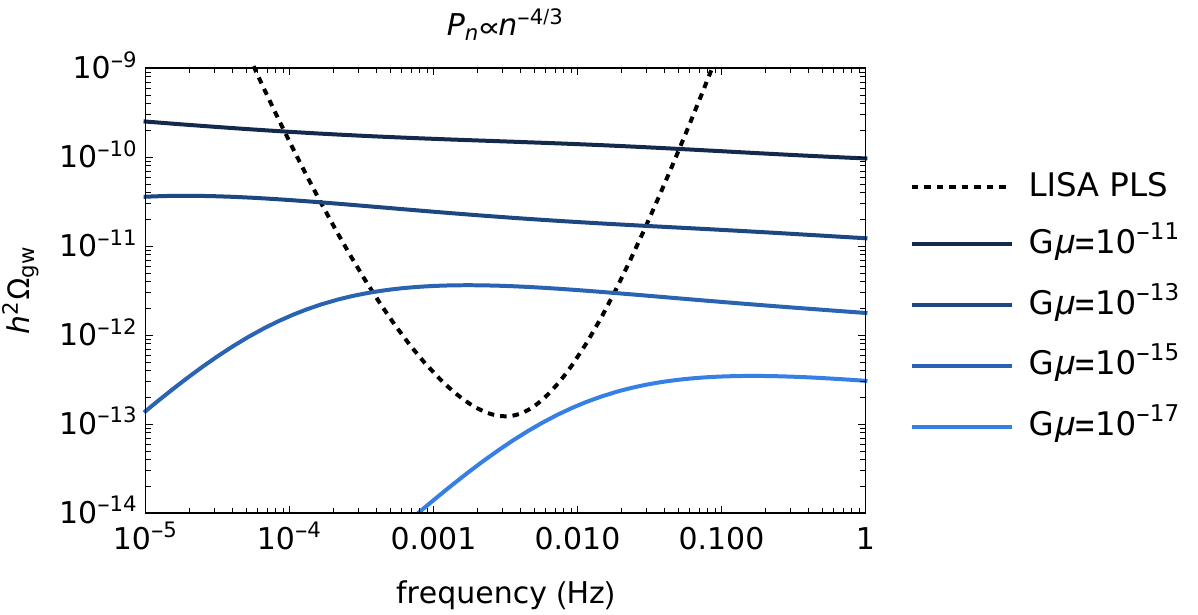}
    \caption{How the SGWB of a loop network (solid blue curves) shifts through the LISA sensitivity band (dotted black curve) as the string tension varies. We assume $P_n\propto n^{-4/3}$ for this figure, but other spectra exhibit similar behavior.}
    \label{fig:LISA-cascade-43}
\end{figure}

The LISA PLS band is well-positioned to set strong constraints on the string tension, due to how the ``bump'' in the SGWB shifts as $G\mu$ decreases. This effect can be seen in \ref{fig:LISA-cascade-43}, where we show how the SGWB curve for a network, assuming $P_n\propto n^{-4/3}$, shifts through the LISA band for varying tension.

\begin{figure}
    \centering
    \includegraphics[scale=1.35]{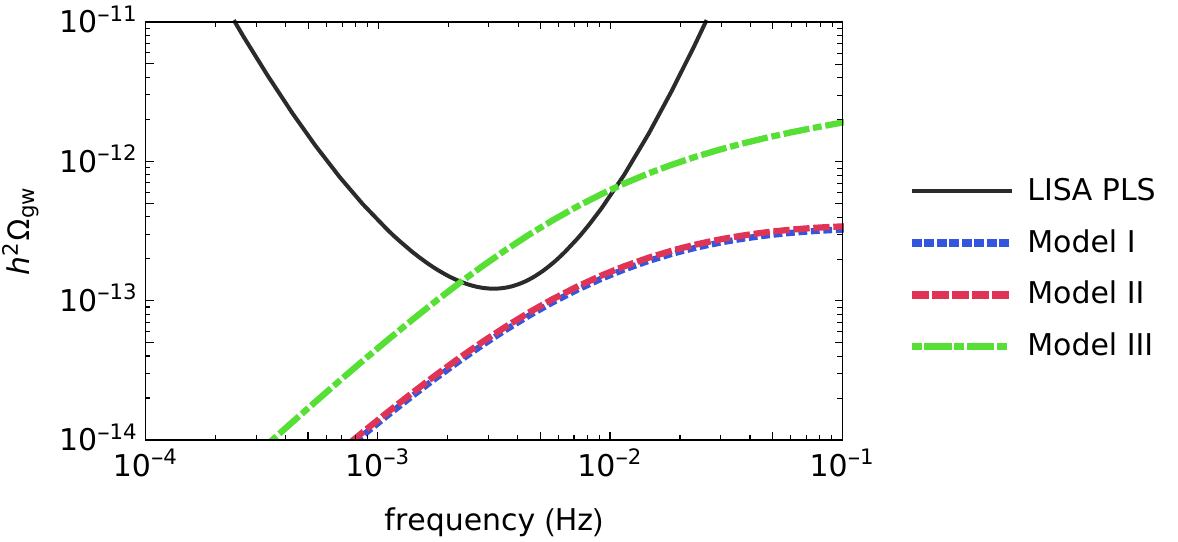}
    \caption{A comparison of the LISA sensitivity curve to the SGWB predicted by all three models using $G\mu=10^{-17}$, $P_n\propto n^{-4/3}$. Models I and II are effectively identical in this regime, due to $\alpha\gg\mathrm{\Gamma} G\mu$. We therefore see that we expect that LISA could only constrain string tensions higher than $G\mu\approx 10^{-17}$.}
    \label{fig:LISA-bounds}
\end{figure}

This shows us that it is the trailing edge of the SGWB bump which will be the last part of that curve to pass through the LISA sensitivity band. By varying the string tension, it is possible to find the lowest $G\mu$ for which this intersection still takes place. While the exact bound depends on our choice of model and $P_n$, in the regime LISA will probe, all three models predict a string tension bound of $\mathcal{O}(10^{-17})$. This is shown in figure~\ref{fig:LISA-bounds}. We chose $P_n\propto n^{-4/3}$ for purposes of comparison, as this is the chromatic index of pure cusps, which are expected to dominate at high frequencies. Other choices include $P_n\propto n^{-5/3}$ (for kinks), $P_n\propto n^{-2}$ (for kink-kink collisions), or an averaged spectrum of loops taken from simulation (cf. figure~\ref{fig:fiducialSGWB}). However, these changes have at most $\mathcal{O}(1)$ effects on the bounds set by LISA.

By comparing eq.~(\ref{eqn:VOSrad}) to eq.~(\ref{eqn:BOSrad}), we see that with our choice of $\alpha$ and $A_r$, these two expressions converge when $\alpha\gg\mathrm{\Gamma} G\mu$. As this is the case here, the curves for Model I and Model II in figure~\ref{fig:LISA-bounds} are effectively identical.

While we are primarily concerned with setting bounds on string tension, it is worth noting here that for string tensions larger than the lower bounds, particularly those of an order of magnitude or more larger, LISA will probe the high-frequency side of the SGWB bump. The particular shape of this region depends on how the degrees of freedom change across the Universe's history. This is additionally important because while the three models all predict roughly equal bounds for the LISA window at this particular tension, Models I and II disagree with Model III at high frequencies. E.g., when $G\mu=10^{-17}$, the plateau for Models I and II happens at $h^2\mathrm{\Omega}_{\rm gw}\approx 6.04\times 10^{-14}$, while Model III's plateau is at $h^2\mathrm{\Omega}_{\rm gw}\approx9.98\times 10^{-9}$. Thus, if strings with a tension much greater than $\mathcal{O}(10^{-17})$ exist, these discrepant regions will pass through the LISA band.

\subsection{Agnostic approach to loop size and intercommutation probability}
\label{ss:inter}

In previous sections, we discussed the results obtained from the largest and more recent Nambu-Goto simulations. In this section, we take a different approach: an ``agnostic'' approach that extends our analysis further by studying the capability of LISA to probe scenarios characterized by different loop sizes $\alpha$ parametrically using Model I. This not only allows us to fully characterize the parameter space available for exploration with LISA, but also to understand LISA's ability to detect string models that deviate from the standard Nambu-Goto scenario. Throughout this section, we will take the normalizing parameter introduced in section \ref{sec:VOS}, $\mathcal{F}=1$ and $f_r=\sqrt{2}$.

\subsubsection{Loop size} \label{sssec:agnostic_loop_size}

Although the typical shape of the SGWB generated by cosmic string networks is roughly independent of $\alpha$, the amplitude of the radiation-era plateau and the height, broadness and location of the peak of the spectrum are determined by the size of the loops that are created (as well as by cosmic string tension). In reality, the amplitude of the spectra generally decreases with decreasing $\alpha$ and, therefore, one would expect LISA to be less sensitive in general to scenarios in which loops are created with a smaller size. In fact, one finds, using eq.~(\ref{eqn:theone}), eq.~(\ref{eqn:energyI}) and eq.~(\ref{eqn:VOSrad}), that the amplitude of the radiation era plateau is given by
\be
\label{eqn:plateaualpha}
\mathrm{\Omega}^{\rm plateau}_{\rm gw}\, h^2=\frac{128}{9}\pi A_r \mathrm{\Omega}_{\rm rad}\,h^2 \frac{G\mu}{\epsilon}\left[\left(\epsilon+1\right)^{3/2}-1\right]\simeq 1.02\times 10^{-2} \frac{G\mu}{\epsilon}\left[\left(\epsilon+1\right)^{3/2}-1\right]\,,
\ee 
where $\epsilon=\alpha/(\mathrm{\Gamma} G\mu)$.

To analyze the capability of LISA to probe scenarios with different loop sizes, we consider two different regimes. Let us first consider the case in which the physical length of loops is, at the time of production, significantly larger than the gravitational backreaction scale, with $\epsilon \gg 1$ (which we shall refer to as large loops, for simplicity). In this case (particularly in the frequency range probed by LISA) the dominant contribution to the SGWB comes, in general, from loops created in the radiation era. As a result, we have roughly $\mathrm{\Omega}_{\rm gw} \propto \alpha^{1/2}$ for $\alpha \gg \mathrm{\Gamma} G\mu$ and fixed $G\mu$ (as eq.~(\ref{eqn:VOSrad}) shows). Indeed, we see by using eq.~(\ref{eqn:plateaualpha}), that for $\epsilon\gg 1$ the amplitude of the radiation era plateau\footnote{Note that this has the same dependence on $\mathrm{\Gamma}$ and $G\mu$ as eq.~(\ref{eqn:plateau-model1}) and, by setting $\alpha=0.1$ and $\mathcal{F}=0.1$, one approximately recovers the result therein.} is given by
\begin{equation}\label{eqn:sca-alpha}
\mathrm{\Omega}^{\rm plateau}_{\rm gw} \, h^2 \simeq 1.02\times 10^{-2}\sqrt{\frac{G\mu \alpha}{\mathrm{\Gamma}}}\,.
\end{equation}

\begin{figure}
\centering
\includegraphics[width=5.5in]{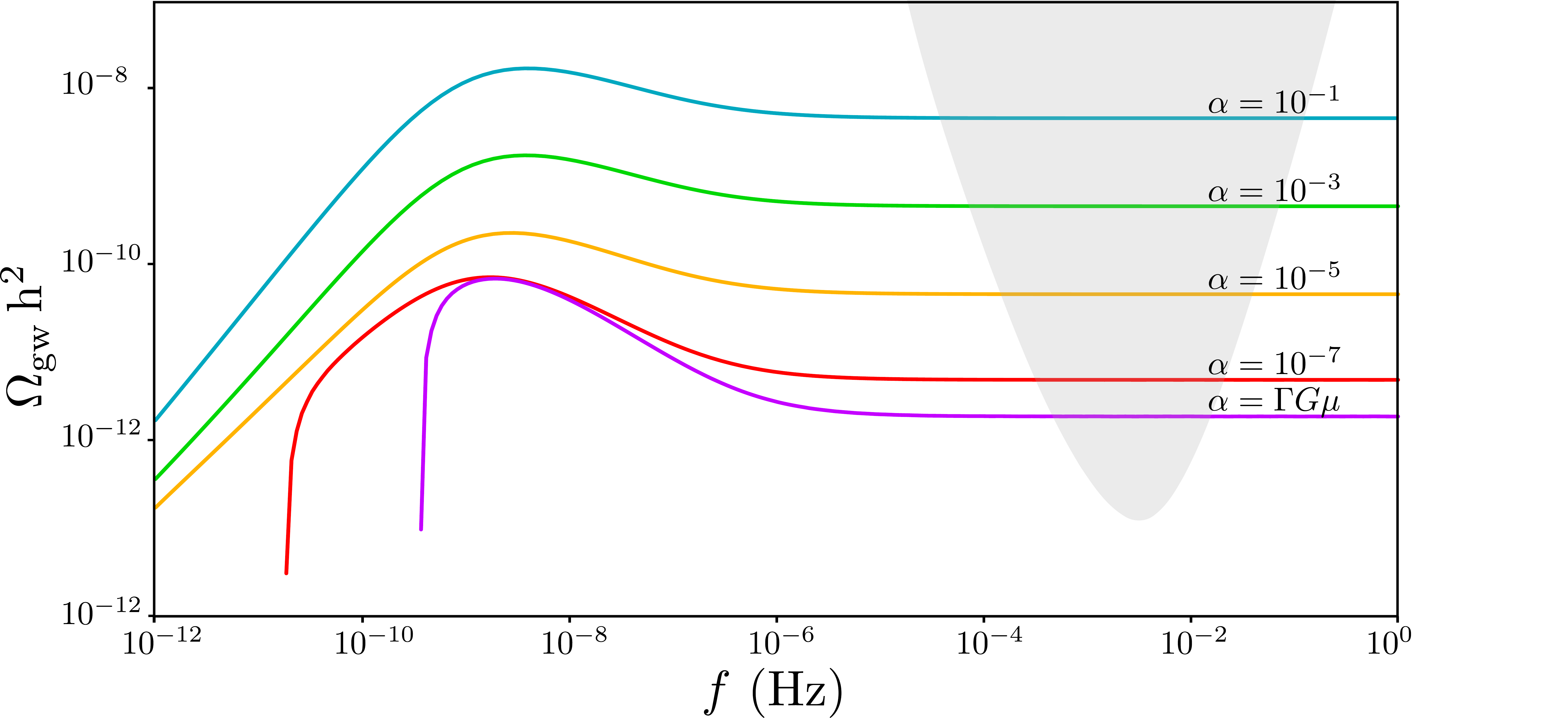}
\caption{The stochastic gravitational wave background generated by cosmic string networks with $G\mu=10^{-10}$ and different values of the loop-size parameter $\alpha$. The shaded area represents the LISA sensitivity window. In these plots, we consider only the fundamental mode of emission and we did not include the change in the effective number of degrees of freedom.}
\label{fig:largealpha1}
\end{figure}

This effect is seen in figure~\ref{fig:largealpha1}, where the SGWB spectra generated by cosmic string networks with $G\mu=10^{-10}$ and different values of $\alpha$ are plotted. Note however that one does not have a mere overall decrease of the amplitude of the spectrum as $\alpha$ decreases. As this figure illustrates, the broadness of the peak of the spectra also decreases as a result of the decrease of the size of loops, since for smaller $\alpha$ loops survive (and emit gravitational waves) for a shorter period of time. We also note that although the relation in eq.~(\ref{eqn:sca-alpha}) is exact for the radiation-era plateau while $\alpha\gg\mathrm{\Gamma} G\mu$, the decrease in the height of the peak starts to slow down as we decrease $\alpha$. This happens due to the fact that, as $\alpha$ decreases and the lifetime of loops is shortened, the number of loops created in the radiation era that decay during the matter era also diminishes. Thus, for sufficiently small $\alpha$, the dominant contribution to the peak of the spectrum are loops produced in the matter era (during which $n(l,t)$ is roughly independent of $\alpha$ for $\alpha\gg\mathrm{\Gamma} G\mu$, as eq.~(\ref{eqn:VOSmat}) shows). As a result, the relative height of the peak of the spectrum in relation to the radiation-era flat region increases as loop size decreases. This also means that, as $\alpha$ is lowered, the effect of the radiation-to-matter transition on the shape and amplitude of the spectra becomes increasingly relevant. As a matter of fact, assuming that cosmic string networks remain in a linear scaling regime after the onset of the radiation-matter transition leads to a significant underestimation of the size and number density of loops produced during the matter era~\cite{Sousa:2013aaa}. On the other hand, as we have seen in section~\ref{ssec:rmtrans}, for $\alpha=10^{-1}$ the effect of this assumption of linear scaling is only observed at frequencies that are outside of the LISA sensitivity window. As we consider smaller loops the effect of the radiation-to-matter transition becomes relevant for the LISA mission. For this reason, we take this effect into consideration in this agnostic forecast of the LISA projected constraints. 

\begin{figure}
    \centering
    \includegraphics[width=5.5in]{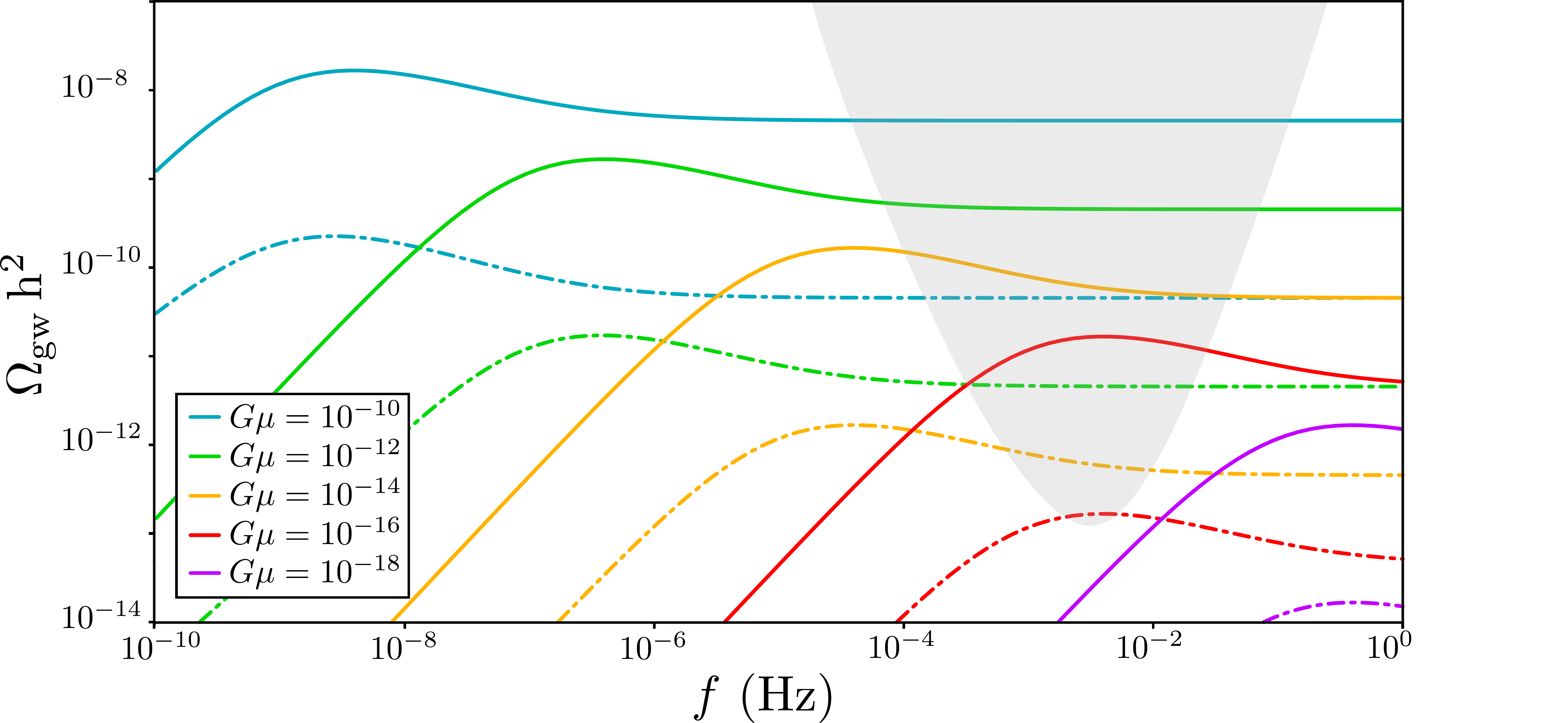}
    \caption{The stochastic gravitational wave background generated by cosmic string networks with $\alpha=10^{-1}$ (solid lines) and $\alpha=10^{-5}$ (dash-dotted lines) for different values of $G\mu$. The shaded area represents the LISA sensitivity window. In these plots, we consider only the fundamental mode of emission and we did not include the change in the effective number of degrees of freedom.}
    \label{fig:largealpha2}
\end{figure}

Another effect that we have to take into consideration when analyzing the sensitivity of LISA to scenarios with different loop sizes is the change of the location of the peak of the spectrum with the variation of $G\mu$. The peak frequency scales approximately as~\cite{Sanidas:2012ee}\footnote{Although this relation was fitted using a framework based on the one-scale model (in which cosmic strings are assumed to be in the linear scaling regime throughout their evolution), we have verified that it still provides a reasonably good approximation, with only small deviations, within the framework we use here.}
\begin{equation}\label{eqn:peakfreq}
f_{\rm peak}\sim \frac{1}{\alpha}\left(2+\frac{\alpha}{\mathrm{\Gamma} G\mu}\right)^{10/9}\,,
\end{equation}
which gives
\begin{equation}
 f_{\rm peak}\propto \alpha^{1/9}\left(\mathrm{\Gamma} G\mu\right)^{-10/9}
\end{equation}
in the large loop regime. For fixed $G\mu$, the dependence on $\alpha$ is weak and the peak appears at approximately the same frequency as shown in figure~\ref{fig:largealpha1} (wherein one can also see that the slight shift towards higher frequencies predicted in eq.~(\ref{eqn:peakfreq}) is present). However, the frequency in which the peak appears depends more strongly on cosmic string tension and, as a result, the peak of the spectrum,  which has a significantly higher amplitude, is expected to shift towards higher frequencies --- and into the LISA window --- as $G\mu$ is lowered. This effect may be seen in figure~\ref{fig:largealpha2}, where we plot the SGWB spectra generated by cosmic string networks with two different values of loop-size parameter $\alpha$ for different values of $G\mu$.

\begin{figure}
    \centering
    \includegraphics[width=5.5in]{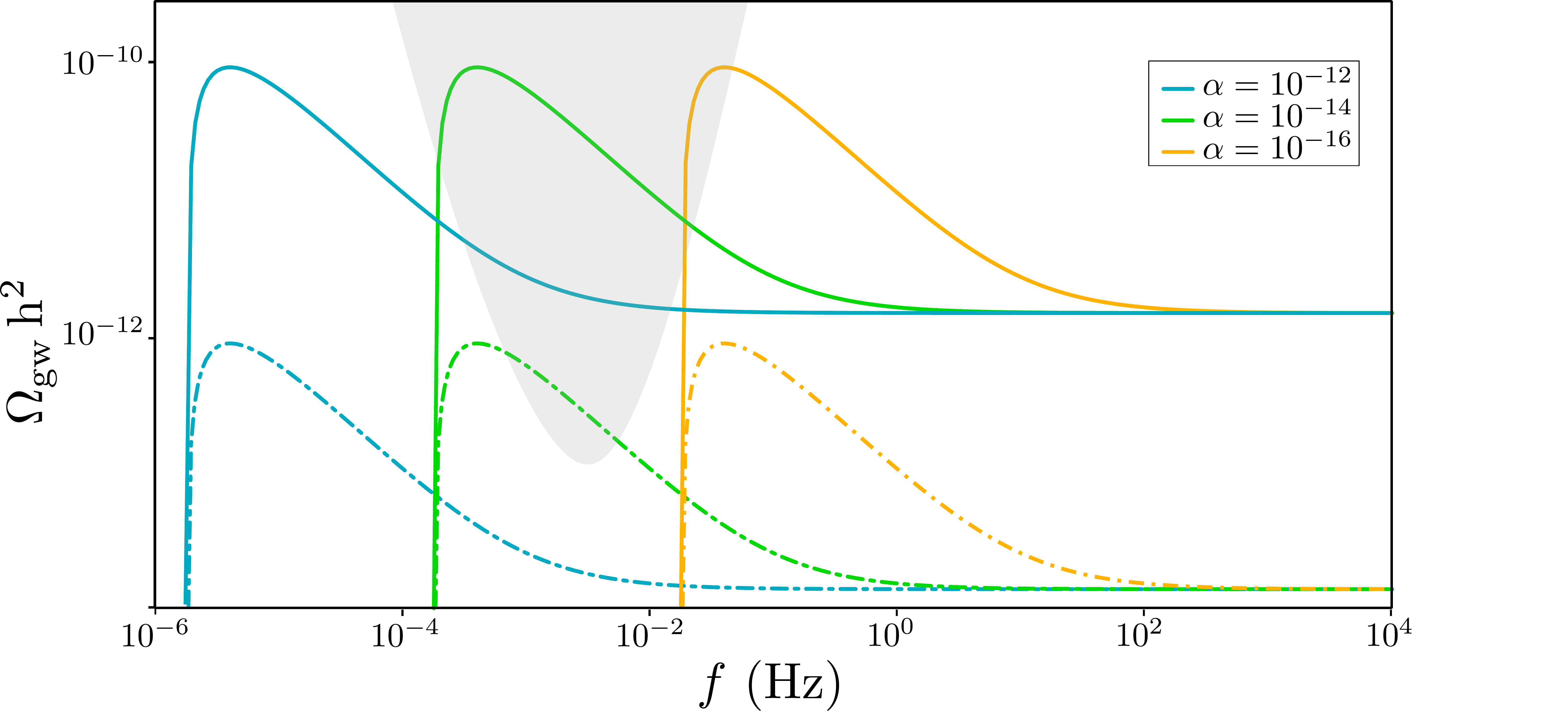}
    \caption{The stochastic gravitational wave background generated by cosmic string networks with $G\mu=10^{-10}$ (solid lines) and $G\mu=10^{-12}$ (dash-dotted lines) for different values of the loop-size parameter $\alpha$ in the small-loop regime. The shaded area represents the LISA sensitivity window. In these plots, we consider only the fundamental mode of emission and we did not include the change in the effective number of degrees of freedom.}
    \label{fig:smallalpha}
\end{figure}

In the small loop regime --- in which the physical length of loops is significantly smaller than the gravitational backreaction scale, with $\alpha\ll \mathrm{\Gamma} G\mu$ --- the shape of the SGWB spectrum is not affected by varying $\alpha$ or $G\mu$. As matter of fact, in this regime, loops survive significantly less than a Hubble time and may, therefore, be regarded as decaying effectively immediately (on cosmological timescales) once they are formed~\cite{Sousa:2014gka}. Thus, a decrease in the size of loops in this regime merely results in a linear shift of the spectrum towards higher frequencies, without any impact on its overall shape. For the same reason, decreasing the value of cosmic string tension merely causes a decrease of the amplitude of the spectrum: $\mathrm{\Omega}_{\rm gw} \propto G\mu$, for fixed $\alpha$. In fact, using  eq.~(\ref{eqn:plateaualpha}), one finds that for $\epsilon\ll 1$ the amplitude of the radiation era plateau is, in this case,
\begin{equation}\label{eqn:plateausmall}
\mathrm{\Omega}^{\rm plateau}_{\rm gw} h^2=\frac{64}{3}\pi A_r h^2\mathrm{\Omega}_{\rm rad}G\mu \simeq 1.52\times 10^{-2} G\mu \,.
\end{equation}
This does not depend on the size of loops $\alpha$ and on $\mathrm{\Gamma}$ and it may, therefore, be regarded as the ``minimal'' amplitude of the radiation era plateau for fixed $G\mu$. This is illustrated in figure~\ref{fig:smallalpha}, where the spectra generated by small cosmic string loops is plotted for different values of $\alpha$ and $G\mu$.

\begin{figure}
    \centering
    \includegraphics[width=5.5in]{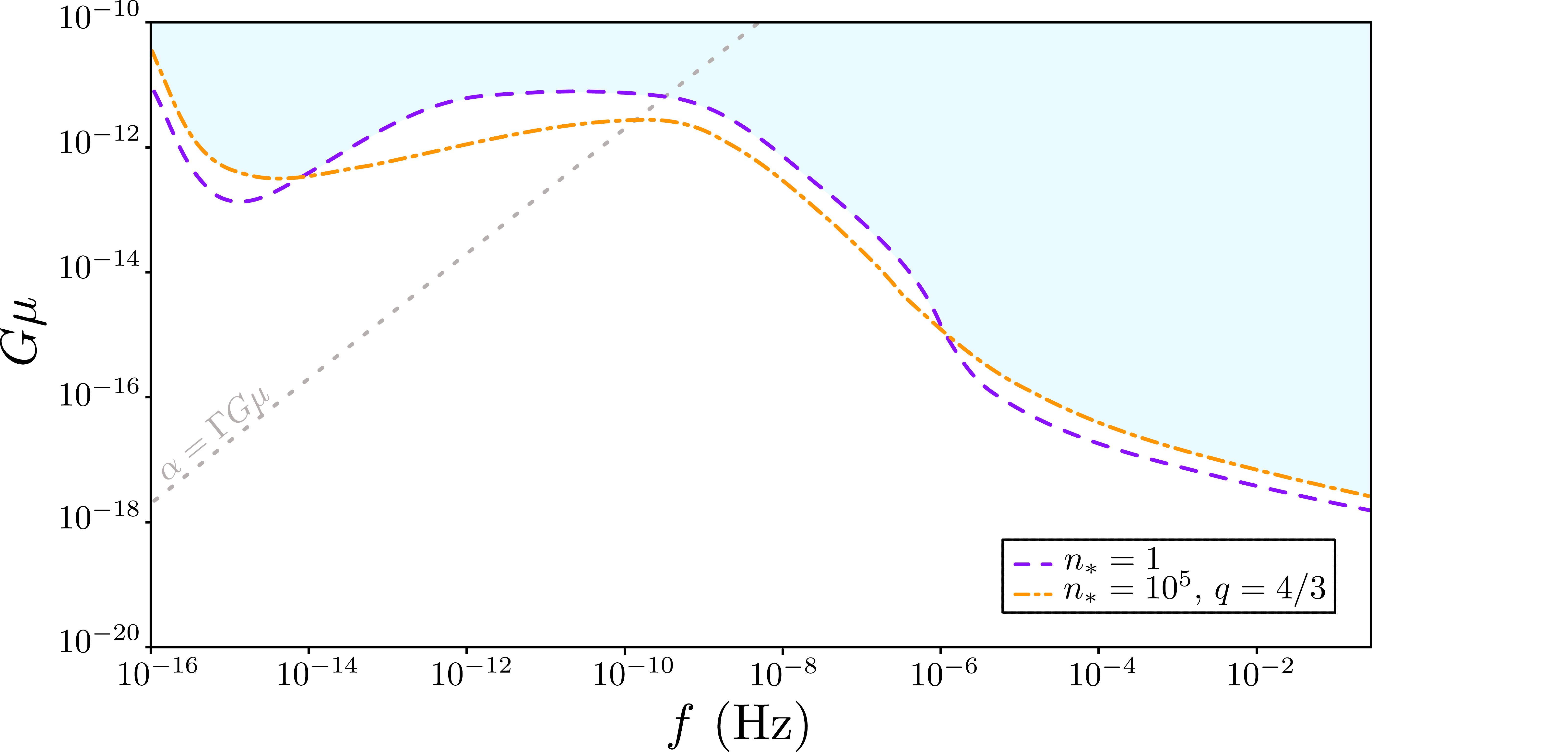}
    \caption{Projected constraints on $G\mu$ of the LISA mission for cosmic string scenarios characterized by different loop-size parameter $\alpha$ for $n_*=1$ (dashed line) and $n_*=10^5$, with $q=4/3$ (dash-dotted line). The shaded area corresponds to the region of the $(\alpha,G\mu)$ parameter space that will be fully available for exploration with LISA. The dotted line corresponds to scenarios for which $\alpha=\Gamma G\mu$, so that the region above this line corresponds to cosmic string models in which loops are small, while the region bellow corresponds to the large loop regime.}
    \label{fig:constraints}
\end{figure}

The combination of all these different effects makes it non-trivial to extend the forecasts computed for a single value of $\alpha$ to significantly different loop sizes. For instance, as figure~\ref{fig:largealpha2} shows, LISA may probe cosmic string networks with $\alpha=10^{-1}$ up to tensions just above $G\mu=10^{-18}$. In the case of networks with $\alpha=10^{-5}$,  however, the maximum tension that LISA will be able to detect is below $G\mu=10^{-16}$, which is significantly lower than one would na\"ively expect from eq.~(\ref{eqn:sca-alpha}). Moreover, figure~\ref{fig:smallalpha} demonstrates that there is a range of $\alpha$ in the small loop regime for which the peak of the spectrum --- which is quite prominent in this regime --- coincides with the LISA window (for some values of $G\mu$) and, therefore, such scenarios may be more strongly constrained with LISA than other scenarios in which $\alpha$ is larger. To take these effects into account, we have performed a numerical computation of the $(\alpha,G\mu)$ parameter space available for exploration with LISA. The results are plotted in figure~\ref{fig:constraints}, and they show us the capability of LISA to probe different cosmic string scenarios characterized by the production of loops with different sizes. Here, we follow the approach introduced in ref.~\cite{Sanidas:2012ee} and present constraints for $n_*=1$ (dashed line) and $n_*=10^5$ (dash-dotted line), where $n_*$ represents the maximum mode of emission included in the simple gravitational wave power spectrum from loops with $q=4/3$.  These curves represent the lowest possible values of the string tension that LISA will be able to probe for each value of $\alpha$, in these two scenarios.

To analyze these results, let us start by considering the small-loop regime. LISA cannot probe the SGWB generated by cosmic string loops to arbitrarily small $\alpha$. This is a mere consequence of the fact that LISA shall only probe a finite frequency window and of the fact that, as we have seen, lowering $\alpha$ in the small-loop regime moves the spectrum towards higher frequencies. As a matter of fact, the minimum frequency emitted by a cosmic string network is that of loops created at the present time, $f_{\rm min}\sim 2/(\alpha t_0)$, and therefore the minimum loop-size parameter that can be probed with LISA is given by
\begin{equation}\label{eqn:alphamin}
\alpha_{min}=6.8\times 10 ^{-18}\,,
\end{equation}
independently of $G\mu$. As a result, scenarios in which the networks produce tiny loops will be beyond the reach of LISA.\footnote{These scenarios are not particularly well physically motivated, since one generally expects smoothing to occur on scales smaller than the gravitational backreaction scale. Nevertheless, several works have reported the existence of such tiny loops~\cite{Vincent:1996qr,Vincent:1997cx}.} In any case, this shows us that, in principle, LISA shall be able to probe cosmic string scenarios spanning about 17 orders of magnitude in loop size.\footnote{Note however that, for tensions compatible with current CMB bounds, the SGWB spectrum ``leaves'' the LISA window for larger values of $\alpha$, around $\alpha \sim 10^{-16}$ (cf. figure~\ref{fig:constraints}).} In the small-loop regime, the amplitude of the peak of the spectrum --- located at $f_{\rm peak}=2\times 10^{-17}/\alpha$ (Hz) --- is given by~\cite{Sousa:2014gka}\footnote{Here, we have included the effect of the redshifting of the peculiar velocities of loops that was not taken into account in the analytical approximation for the SGWB spectrum generated by small loops in ref.~\cite{Sousa:2014gka}.}
\begin{equation}\label{eqn:peaksmall}
\mathrm{\Omega}_{\rm peak}h^2\simeq 60  \mathrm{\Omega}^{\rm plateau}_\text{\rm gw}h^2 \simeq 9.1\times 10^{-1}G\mu\,.
\end{equation} 
One then finds that LISA will not be able to probe small-loop models for $G\mu<1.3\times 10^{-13}$. This is, thus, the most stringent bound that LISA may put on the cosmic string tension in scenarios in which loops are created with small size. This value corresponds to the case in which $\alpha \sim 10^{-15}$ and $n_*=1$, as figure~\ref{fig:constraints} shows, for which the peak of the spectrum coincides with the maximum sensitivity of the LISA window (cf. figure~\ref{fig:smallalpha}). 

In the large-loop regime, as we have seen, the amplitude of the spectra is highly dependent on the size of loops and, for this reason, so is the strength of the constraints that LISA may put on cosmic string tension. The amplitude of the radiation era plateau of spectrum for small loops in eq.~(\ref{eqn:plateausmall}) may be regarded as the ``minimal'' amplitude of this plateau. Thus one may use it to derive the value of cosmic string tension above which all cosmic string scenarios in which loop production is significant (with $\alpha>10^{-16}$) are excluded:
\begin{equation}\label{eqn:safebound}
G\mu<8.0 \times 10^{-12}\,.
\end{equation}
This provides us with the safest (yet most conservative) model-independent LISA bound on cosmic string tension, which corresponds to the value of the plateau observed at the mid-$\alpha$ range in figure~\ref{fig:constraints}. Note however that LISA shall be able to establish significantly more stringent constraints for the largest possible $\alpha$. Indeed, LISA may go seven orders of magnitude beyond the bound in eq.~(\ref{eqn:safebound}) for $\alpha=10^{-1}$:
\begin{equation}
G\mu (\alpha=10^{-1}) <3.4\times 10^{-18}\,.
\end{equation} 
As we have seen in section~\ref{ssec:gmu-constraints}, this corresponds approximately to the case of Nambu-Goto strings, apart from a factor of $\sim 0.1$ (since in that case only about $10\%$ of the energy lost by the network goes into gravitational radiation). Apart from this factor, these results are in agreement with those presented in section~\ref{ssec:gmu-constraints}.

\subsubsection{Intercommutation probability}
\label{subsec:intercomP}

In this paper,  we have assumed that the intercommutation probability $\mathcal{P}$ is equal to 1. In effect, this amounts to stating that when two strings collide, they exchange partners every time. Indeed, in the Abelian-Higgs model (in the BPS and ``type II'' regime), the collision of two straight strings with velocities $\pm v$ and relative orientation given by an angle $\theta$ essentially always (that is, in nearly all of the $(\theta,v)$ parameter space) leads to the strings exchanging partners during the collision (see ref.~\cite{Shellard:1987bv,Verbiest:2011kv}). Setting $\mathcal{P}=1$ is then equivalent to this statement. For other field theory strings, such as Abelian-Higgs strings in the type I regime, the collision may lead to other outcomes, such as the formation of a junction~\cite{Salmi:2007ah,Bevis:2009az}. We do not consider these more complicated cases here.

Recent development in String Theory suggest that fundamental strings (or F-strings) and 1-dimensional Dirichlet branes (or D-strings) may be stretched to macroscopic sizes and play the cosmological role of cosmic superstrings. The copious production of these cosmic superstrings is, in fact, predicted to occur at the end of several brane-inflationary scenarios (see, e.g., refs.~\cite{Dvali:2001fw,Jones:2003da,Copeland:2003bj}).

Cosmic superstrings may have a intercommutation probability $\mathcal{P}$ significantly smaller than unity, as a result of their quantum nature: in fact it has been shown~\cite{Jackson:2004zg} that $10^{-3} \lesssim \mathcal{P} \lesssim 1$ in collisions between F-strings and $10^{-1} \lesssim \mathcal{P} \lesssim 1$ for D-string collisions. When a FF- or DD-string collision occurs, the strings may then --- unlike ordinary strings --- pass through each other without intercommutation. For this reason, cosmic superstring networks are expected to lose energy less efficiently. Their energy density, and consequently the amplitude of the SGWB they generate,  may therefore be expected to be larger than that of ordinary strings. Hence the constraints derived on $G\mu$ in this paper are  conservative: with $\mathcal{P}< 1$, the bounds on $G\mu$ will be tighter (see, e.g., ref.~\cite{Aasi:2013vna} for a discussion of this effect at LIGO frequencies). In general, one expects the loop-chopping parameter of these networks to be such that
\be
\cc(\mathcal{P})=\cc(1)\mathcal{P}^\gamma\,,
\ee
where $\cc(1)=\cc=0.23$ is the loop-chopping parameter of ordinary strings (which have $\mathcal{P}=1$). Although one may na\"ively expect, within the one-scale framework, $\gamma=1$~\cite{Jones:2003da}, numerical simulations indicate that this effect is less dramatic due to an accumulation of small-scale structure on cosmic strings with reduced intercommutation probability. It has been observed that $\gamma=1/2$ in Nambu-Goto simulations in Minkowski space~\cite{Sakellariadou:2004wq} and $\gamma=1/3$ in both radiation- and matter-era simulations~\cite{Avgoustidis:2005nv}. Since the exact value of $\gamma$ is still a matter for debate, here we restrict ourselves to a (mostly) qualitative discussion of the effects of $\mathcal{P}$.

Weakly interacting networks, with $\cc \ll 1$, scale in the radiation era according to $\xi=\sqrt{2}\cc$ and $\vv^2\approx 1/2$~\cite{Avelino:2012qy}. Therefore, one may, in general, expect the amplitude of the radiation era plateau of the SGWB to scale as~\cite{Sousa:2016ggw}
\be\label{eqn:lowp-sca}
 \mathrm{\Omega}^{\rm plateau}_\text{\rm gw}\propto \cc^{-2}\propto \mathcal{P}^{-2\gamma}\,.
\ee
Note however that, in this case, the length of the loops created is not known. There is some evidence that the reduction of the intercommuting probability is more efficient in suppressing the production of large loops than of small loops~\cite{Avgoustidis:2005nv}, which seems to indicate that smaller $\alpha$ ($\sim \Gamma G\mu$) may be favoured for these networks. However, the precise number density of loops has not been determined using numerical simulations yet. Nevertheless, one may obtain, using eqs.~(\ref{eqn:safebound},\ref{eqn:lowp-sca}), a conservative $\alpha$-independent constraint on the cosmic string tension of networks with $\mathcal{P}\ll 1$. This bound --- corresponding to the lowest $G\mu$ for which the SGWB is within the reach of LISA for all values of $\alpha$ --- is presented in Table~\ref{tab:lowp} for $\mathcal{P}=10^{-1},10^{-2},10^{-3}$.

Naturally, as with ordinary strings, LISA will impose tighter constraints for scenarios in which $\alpha$ is large. The most stringent constraint on $G\mu$ will necessarily be those of scenarios characterized by the largest possible $\alpha$, with $\alpha_{\rm max}\sim 0.3 \mathcal{P}^\gamma$ (corresponding to the characteristic length of the network which may, in this case, be significantly smaller than the horizon). These constraints are also recorded in Table~\ref{tab:lowp} for the same values of $\mathcal{P}$. These two constraints are then indicative of the ability of LISA to detect cosmic string scenarios with a reduced intercommutation probability. 

\begin{table}
\begin{center}
\begin{tabular}{c|c|c|c|c|}
\cline{2-5}
                                    & \multicolumn{2}{c|}{Conservative}           & \multicolumn{2}{c|}{Stringent}                \\ \hline
\multicolumn{1}{|l|}{$\mathcal{P}$} & $\gamma=1/2$         & $\gamma=1/3$         & $\gamma=1/2$         & $\gamma=1/3$         \\ \hline
\multicolumn{1}{|l|}{$10^{-1}$}     & $8.0\times 10^{-13}$ & $1.7\times 10^{-12}$ & $3.4\times 10^{-19}$ & $5.6\times 10^{-19}$ \\ \hline
\multicolumn{1}{|l|}{$10^{-2}$}     & $8.0\times 10^{-14}$ & $3.7\times 10^{-13}$ & $2.9\times 10^{-20}$ & $1.2\times 10^{-19}$ \\ \hline
\multicolumn{1}{|l|}{$10^{-3}$}     & $8.0\times 10^{-15}$ & $8.0\times 10^{-14}$ & $8.5\times 10^{-21}$ & $2.9\times 10^{-20}$ \\ \hline
\end{tabular}
\end{center}
\caption{Projected constraints of the LISA mission for cosmic string scenarios with reduced intercommutation probability $\mathcal{P}$. Here, ``Conservative'' refer to the safe ($\alpha$-independent) bounds obtained using the minimal amplitude of the radiation-era plateau, while the constraints labeled as ``Stringent'' correspond to those of scenarios with the largest possible $\alpha$.}
\label{tab:lowp}
\end{table}

However, we note that there are relevant aspects of cosmic superstring dynamics that were not taken into account when deriving these constraints. In particular, when superstrings of different types collide, they are expected to bind together to create a third type of string, which has a higher tension than its two constituents. This is expected to lead to networks with junctions and a hierarchy of tensions~\cite{Copeland:2003bj}. The creation of junctions is expected to have an impact on the large scale dynamics of cosmic string networks~\cite{Copeland:2006eh,Copeland:2006if,Copeland:2007nv,Avgoustidis:2007aa,Rajantie:2007hp,Sakellariadou:2008ay,Avgoustidis:2009ke,Avgoustidis:2014rqa} and therefore to affect the shape and amplitude of the SGWB generated by cosmic superstrings~\cite{Pourtsidou:2010gu,Sousa:2016ggw}. Moreover,  there are several other important aspects regarding the gravitational wave emission by cosmic superstrings that need to be clarified --- most notably the number and strength of the cusps \cite{Elghozi:2014kya} as well as the possible coupling of superstrings to other fields --- before a detailed study of the parameter space available to LISA can be performed.

\subsection{Gravitational wave bispectrum from long strings}\label{ssec:BisectrumInfiniteStrings}

The GW signal due to the gravitational decay of loops that we have analyzed in this paper cannot be resolved beyond its stochastic nature, and it is expected to be Gaussian.\footnote{In reality, on top of the continuous stochastic Gaussian background from cosmic strings, there can be individual bursts emitted by nearby strings or a ``popcorn'' discontinuous noise~\cite{Regimbau:2011bm}; recall the discussion in section~\ref{ssec:popcorn}. These signals due to bursts represent, in a sense, a temporal deviation from Gaussianity, that can be measured from the two-point function. However, they do not correspond to the type of non-Gaussianity that we are referring to in this subsection, as they do not form a continuous stochastic background.} The irreducible emission of GWs from a defect network (described in section~\ref{ssec:AH-FT}) is however expected to be highly non-Gaussian. This is simply due to the fact that the source of the GWs is bilinear in the amplitude (modulo derivatives) of the fields of which the cosmic strings are made. This implies that any correlator of an odd number of tensor perturbations will be characterized by the correlation of an even product of fields, which is non-vanishing even if the fields were Gaussian. We therefore expect that any non-Gaussianity in the continuous stochastic background sourced by a cosmic string network is due to the irreducible GW emission, even if this signal is sub-dominant in terms of amplitude of power spectra.

The capability of LISA to detect 3-point correlations of SGWBs has been recently analyzed in detail in ref.~\cite{Bartolo:2018qqn}. At present, the 3-point GW function of this background can be estimated analytically only in a simplified case, namely in the large-$N$ limit of a global phase transition due to the spontaneous symmetry breaking of $O(N)$ into $O(N-1)$. The GW background due to the dynamics of such global defects has been estimated in the limit $N \gg 1$ (see section\ref{ssec:InfiniteStrings} and in particular eq.~(\ref{eq:GWSOSFplateau})). The 3-point function (in the equilateral configuration) has been presented in ref.~\cite{Adshead:2009bz}. Order of magnitude calculations in the large-$N$ limit leads to a GW bispectrum peaked in the equilateral configuration as~\cite{Adshead:2009bz}
\begin{equation}\label{eqn:NGglobal}
    k^6\mathcal{B}(k,k,k) \sim C_{\rm NL}(k^{3/2}\mathcal{P}_h(k))^2\qquad\mathrm{with}\qquad C_{\rm NL} \sim \frac{3.6}{\sqrt{N}}\,,
\end{equation}
where $\mathcal{P}_h$ is the total power spectrum (summing over the two polarizations) and $N \gg 1$ is the number of components of the symmetry-breaking field. This is of course a very rough estimate for global strings, for which  $N = 2$, and we certainly do not know how this relation is modified in the case of Abelian-Higgs strings. However, eq.~(\ref{eqn:NGglobal}) suggests very clearly that, in general, that we should expect a large departure from Gaussianity for the irreducible GW background from any defect network.

Let us note that even though ref.~\cite{Bartolo:2018qqn} has provided a formalism to characterize a potential detection by LISA of the bispectrum of a SGWB, refs.~\cite{Bartolo:2018evs,Bartolo:2018rku} have recently pointed out that propagation effects of GWs across a perturbed universe like ours --- from the generation point to the LISA detector --- will suppress the bispectrum to  un-observable levels. This suppression is expected to be present for any non-Gaussian SGWB, as long as the signal consists of GWs that have traveled across cosmological scales. If this claim is finally sustained, it will essentially imply that independently of the level of (intrinsic) non-Gaussianity of a given SGWB, the 3-point function of GWs will be never measured by direct detection detectors\footnote{Note that measurements of a 3-point function of perturbations in the CMB evade this problem.}. 

\section{Discussion and conclusion}\label{sec:overview}

In this paper we have analyzed the ability of LISA to detect and characterize a SGWB produced by a network of cosmic strings. Our key finding is that LISA will be able to probe cosmic string with tensions $G\mu\gtrsim\mathcal{O}(10^{-17})$, under a "standard" set of assumptions: namely, that the string dynamics are accurately described by the Nambu-Goto action, that colliding strings always intercommute, and that the average loop size at formation (in units of cosmic time $t$) is $\alpha\approx 0.1$. This presents an improvement of $\sim 6$ orders of magnitude over current constraints from pulsar timing arrays (PTA), and potentially in $\sim 3$ orders of magnitude over estimated future constraints from next generation of PTA experiments\footnote{The reason for this is that the spectrum of the SGWB from cosmic strings, shifts towards larger frequencies for small tensions, and hence 'leaves behind' the frequency window accessible to PTA experiments, no matter how precise these may become. See figures~\ref{fig:fiducialSGWB} and~\ref{fig:combined-LRS-large}.}. We have also explored scenarios in which the latter two assumptions are relaxed.
Decreasing the loop size at formation $\alpha$ generically leads to weaker constraints on the string tension.
Decreasing the intercommutation probability $\mathcal{P}$ leads to a range of possible constraints, due to the uncertainty about $\alpha$ in these scenarios; however, for larger values of $\alpha$, networks with a small intercommutation probability are very strongly constrained, with LISA being able to reach tensions as small as $G\mu\approx10^{-20}$.

In addition, we have discussed how a detection of the string SGWB could be used to probe fundamental physics, such as changes in the number of relativistic degrees of freedom, or the inclusion of transient epochs prior of radiation domination, characterized by a non-standard equation of state. Such studies are of particular interest for LISA, because its detection window is well-positioned to measure the segment of the string SGWB which contains information about these processes (in the event that $G\mu$ is at least an order of magnitude above the lower bound). Thus, a detection of cosmic strings is of use and interest to the cosmology and particle physics communities at large.

Note that in our assessment of detection we have assumed an ideal case in which the stream data to be measured by LISA is perfectly cleaned from all resolvable sources, glitches, and any impurities in general. In particular, we assume that the presence of the foreground of galactic binaries can be subtracted exploiting its yearly modulation~\cite{Adams:2013qma}. We consider the only signal on top of LISA's intrinsic noise to be that of the homogeneous and isotropic stochastic GW background from the sub-horizon loops of a string network. Future work will quantify the ability of LISA to reconstruct the spectral shape of the SGWB for the lowest tensions that can be probed, as well as possible spectral features due to changes in the number of degrees of freedom. For this we plan to use the recent technique for systematic reconstruction SGWB signals without assuming any specific spectral template~\cite{Caprini:2019pxz}. 

Finally, we remark that we have not discussed the GW signal from Abelian-Higgs simulations, nor considered how the dynamics of cosmic superstring networks would alter LISA's detection prospects. Most importantly, no simulation to date has included the real effect of back-reaction on the string network\footnote{See however Refs.~\cite{Quashnock:1990wv,Wachter:2016rwc,Blanco-Pillado:2018ael,Chernoff:2018evo,Blanco-Pillado:2019nto,Helfer:2018qgv} for work along this direction.} (i.e., gravitational self-interaction), and therefore the best that can be done is to model back-reaction with some ansatz. Our results are therefore predicated on such ansatzs representing good approximations to how true back-reaction would affect the SGWB from a string network.

\vspace*{0.3cm} {\it Acknowledgements}. ACJ is supported by King’s College London through a Graduate Teaching Scholarship. M.S. and M.L. are supported in part by the Sci-ence and Technology Facility Council (STFC), United Kingdom, under the research grant ST/P000258/1. M.L.~is also partly supported by the Polish National Science Center grant 2018/31/D/ST2/02048. L. S. is supported by FCT - Funda\c{c}\~{a}o para a Ci\^{e}ncia e a Tecnologia through contract No. DL 57/2016/CP1364/CT0001. This work was also supported by FCT through national funds (PTDC/FIS-PAR/31938/2017) and by FEDER - Fundo Europeu de Desenvolvimento Regional through COMPETE2020 - Programa Operacional Competitividade e Internacionaliza\c{c}\~{a}o (POCI-01-0145-FEDER-031938) and by FCT/MCTES through national funds (PIDDAC) through grant No. UID/FIS/04434/2019. J.J.B.-P. is supported in part by the Spanish Ministry MINECO grant (FPA2015-64041-C2-1P), the Basque Government grant (IT-979-16) and the MCIU/AEI/FEDER grant (PGC2018-094626-B-C21) as well as by the Basque Foundation for Science (IKERBASQUE). J.M.W. is also supported by the Basque Government grant (IT-979-16). S.K. is partially supported by JSPS KAKENHI No.17K14282 and Career Development Project for Researchers of Allied Universities. DGF was supported by the ERC-AdG-2015 grant 694896 and the Swiss National Science Foundation (SNSF) until August 2019, and by a Ram\'on y Cajal contract by Spanish Ministry MINECO, with Ref.~RYC-2017-23493,  from September 2019 onward. J.J.B.-P. and J.M.W. would like to thank Ken D. Olum for useful conversations. M.S. thanks C. Ringeval for several discussions. D.A.S and P.A. thank C.Ringeval and T.Vachaspati for many useful discussions. 

\bibliographystyle{h-physrev4}
\bibliography{GWstringsLISA}

\appendix

\section{Nambu-Goto dynamics}\label{app:NG}

The dynamics of relativistic zero-thickness strings can be obtained from the Nambu-Goto action (see for example \cite{Vilenkin:2000jqa} and references therein),
\begin{equation}
S_{\rm NG} = - \mu \int{d^2\xi \sqrt{-\gamma} },
\end{equation}
where $\mu$ parametrizes the tension of the string, and the integral describes the area of the string worldsheet, whose induced metric is given by $\gamma$.

The equations of motion from this action can be solved in flat space in the gauge where the most generic solution can be shown to be of the form
\begin{equation}
X^{\mu} (\sigma, t) = \frac{1}{2} \left[ X^{\mu}_{-} (\sigma_{-}) + X^{\mu}_{+} (\sigma_{+})   \right],
\end{equation}
where $\sigma$ and $t$ are spacelike and timelike coordinates, respectively, on the worldsheet, and we have introduced $\sigma_{\pm} = t\pm \sigma$. Furthermore we fix the gauge to $X^0_{\pm}= \sigma_{\pm}$, and the spatial part of these functions are normalized so that $|{\bf{X}}_{\pm}'|=1$.

\subsection{Loop dynamics}
Using the solutions found earlier, one can describe the evolution of a loop in its rest frame with the periodic functions ${\bf{X}}_{\pm} (\sigma_{\pm}) = {\bf{X}}_{\pm} (\sigma_{\pm}+l)$. This implies that 
\begin{equation}
\int_{0}^{l} {\bf{X'}}_{\pm} (\sigma_{\pm})  d\sigma_{\pm} = 0\,,
\end{equation}
where $l$ is the length of the loop. This, together with the unit normalization, means that the functions ${\bf{X'}}_{\pm} (\sigma_{\pm})$ would trace out a loop on the Kibble-Turok sphere whose center of mass is at the center of the sphere. These trajectories will therefore generically cross at points where
\begin{equation}
{\bf{X'}}_{-} (\sigma^c_{-}) = {\bf{X'}}_{+} (\sigma^c_{+})\,.
\end{equation}
These special points in the string evolution are called cusps, and it is easy to check that they
represent instants during the string's periodic motion when the string doubles back onto itself, $d{\bf X}/d\sigma=0$, and therefore moves at the speed of light, $|d{\bf X}/dt|=1$. On the other hand, string intersections can lead to intercommutations, which lead to kinks on both the previously-existing string and the newly-formed loop. Kinks are discontinuities of either of ${\bf{X'}}_{\pm}(\sigma_{\pm})$.

The consequences for GW emission of these two type of features are discussed in~\ref{app:2}.

\subsection{Gravitational wave power from cusps and kinks}
\label{app:2}

Solving the linearised Einstein equation for a single Nambu-Goto cosmic string loop, one can write the GW strain in the local wave zone as a mode sum~\cite{Damour:2001bk}
    \begin{equation}
        \label{eq:gw-strain-mode-sum}
        \bar{h}_{\mu\nu}(t,{\bf x})\approx\frac{4G\mu l}{r}\sum_n\exp\left[-\frac{4\uppi\mathrm{i}n}{l}(t-r)\right]I_{n,+}^{(\mu}I_{n,-}^{\nu)}\,,
    \end{equation}
where $r\equiv|{\bf x}|$ is the distance to the source, and $l$ is the invariant loop length. The motion of the loop worldsheet is parametrized by the functions $X^\mu_\pm(\sigma_\pm)$ and contributes to the GW signal through the integrals
    \begin{equation}
        \label{eq:integral-plus-minus}
        I_{n,\pm}^\mu\equiv \frac{1}{l}\int_0^ld\sigma_\pm\exp\left[-\frac{2\uppi\mathrm{i}n}{l}X_\pm^\mu k_\mu\right]\partial_\pm X_\pm^\mu \,,
    \end{equation}
where $k^\mu=(1,{\bf x}/r)$ is a null wavevector and $\partial_\pm=\partial/\partial\sigma_\pm$.\footnote{In the calculation of the GW loop power spectrum, we define the analogous function in a coordinate system whose $z$ axis is in the ${\bf \hat \Omega}$ direction. In this case we can write ${\bf I}^{\pm}_n= \frac{1}{l} \int_0^ld\sigma_\pm\exp\left[-\frac{2\uppi\mathrm{i}n}{l} \left( \sigma_{\pm} - X_z(\sigma_{\pm}) \right)  \right] {\bf X'}_\pm(\sigma_{\pm})$.
}
The $n=\pm1$ frequencies correspond to the fundamental mode of the loop (set by the period of loop oscillation $T=l/2$),
    \begin{equation}
        f_1\approx\frac{d_H}{l}\times10^{-18}\text{ Hz},
    \end{equation}
which for many loops is far below the LISA frequency window of $10^{-4}$--$10^{-2}\,\text{Hz}$ (unless there were loops of size $l$ many orders of magnitude smaller than the present-day Hubble length $d_H$). We are therefore typically concerned with much higher frequencies $f\gg f_1$, i.e.~very high-order harmonics of the loop, $|n|\gg1$. In this limit, the integrals in eq.~\eqref{eq:integral-plus-minus} are \emph{generically} exponentially suppressed for large $n$, and there is little contribution to the GW signal at high frequencies.

There are two important exceptions where the integrals in eq.~(\ref{eq:integral-plus-minus}) are \emph{not} exponentially suppressed, and have a much weaker power-law decay with frequency: (i) when there is a saddle point in the phase, $k_\mu\partial_\pm X_\pm^\mu=0$; (ii) when the function $\partial_\pm X_\pm^\mu$ is discontinuous.

In order to obtain a GW strain that is not exponentially suppressed, one or the other of these conditions must hold for both sets of integrals $I_{n,+}^\mu$ and $I_{n,-}^\mu$. This gives rise to three possibilities~\cite{Damour:2001bk,Vilenkin:2000jqa,Ringeval:2017eww}:
    \begin{enumerate}
        \item Both sets of integrals have a saddle point in the phase, i.e. there are points $X_\pm^{\mu*}$ such that $k_\mu\partial_\pm X_\pm^{\mu*}=0$. These points are then necessarily the same, $X_+^{\mu*}=X_-^{\mu*}$. Physically, we interpret this as an event where part of the loop moves at the speed of light, forming a sharp, transient feature;  this is what we referred to as a \emph{cusp} earlier. The cusp emits a GW burst, which is beamed along the spatial direction corresponding to $X_\pm^{\mu*}$, with an opening angle  $\theta_\mathrm{b}=[1/(g_2 f l)]^{1/3}\approx[2/(\sqrt{3}n)]^{1/3}$.
        \item One of the sets of integrals has a saddle point, while the other has a discontinuity in the integrand, which is interpreted as a discontinuity in the shape of the loop; this is what we called a \emph{kink} before. In this case, the power-law scaling for $\bar{h}_{\mu\nu}$ occurs not just centred on a single direction (as for a cusp), but around a one-dimensional, ``fan-like'' set of directions. We interpret this as the kink propagating around the loop, beaming GWs as it does so, with the beam opening angle being given again by $\theta_\mathrm{b}=[1/(g_2 f l)]^{1/3}\approx[2/(\sqrt{3}n)]^{1/3}$.
        \item Both sets of integrals have a discontinuity at the same point on the worldsheet. This case corresponds to two kinks, one left-moving and one right-moving, meeting each other. We call this a \emph{kink-kink collision}. In this case, there is no saddle point condition to determine a preferred direction, so the GW emission is isotropic rather than beamed.
    \end{enumerate}
In each of these three cases, one can calculate the asymptotic $|n|\gg1$, $f\gg f_1$ GW waveform, and take the Fourier transform of this to get the strain spectrum $\tilde{h}(f)$. This gives
    \begin{equation}
        \label{eq:cusp-kink-collision-waveforms}
        \tilde{h}_\mathrm{c}(f,{\bf r})= g^c_1\frac{G\mu l^{2/3}}{r f^{4/3}},\qquad
        \tilde{h}_\mathrm{k}(f,{\bf r})= g^k_1\frac{G\mu l^{1/3}}{r f^{5/3}},\qquad
        \tilde{h}_\mathrm{kk}(f,{\bf r})= g^{kk}_1\frac{G\mu}{r f^2},
    \end{equation}
for the cusp, kink, and kink-kink collision cases, respectively, taking care to account for the beaming angle in the cusp and kink cases.

Using these expressions, one can obtain the total power emitted for these events by performing the following integral:
\begin{equation}
P= \frac{1}{T} \int_0^{\infty}{df \frac{\pi f^2}{2G} \int_{S^2} d^2{\bf{r}}~r^2\tilde{h}_i^2 (f,\bf{r})}\,.
\end{equation}
For example, in the case of cusps, one can estimate the power to be
\begin{equation}
P= \frac{3\pi^2 g_1^2}{2^{1/3} g_2^{2/3}} \left(G \mu^2\right)\,,
\end{equation}
so we can say that a typical loop with $N_c$ cusps will have a power of order of
\begin{equation}
\mathrm{\Gamma} =  N_c \frac{3\pi^2 g_1^2}{2^{1/3} g_2^{2/3}}\,.
\end{equation}
This expression allows us to relate the parameters of the cusp waveform $g_1$ and $g_2$ to the total power emitted from this loop when the loop is assumed to emit only in the form of cusps. This relation is important in order to make a consistent calculation of the total SGWB and compare Methods I and II.\textbf{}



\textbf{}


\end{document}